%% file: ms.tex
\documentclass[preprint2]{aastex}
\shorttitle{Radio Image of the R CrA Region}
\shortauthors{Choi et al.}

\usepackage{times}
\newcommand{\Jypb}{{Jy beam$^{-1}$}}
\usepackage{epsfig}
\newcommand{\ploteps}[2]{\centerline{\ifdim#2=0mm\epsfig{file=#1}%
                                     \else\epsfig{file=#1,width=#2}\fi}}
\newcounter{panel}[figure]
\renewcommand{\thepanel}
             {{\count20=1\advance\count20by\value{figure}%
               \stepcounter{panel}\small%
               \sc Fig. \the\count20\it\alph{panel}}}
\newcommand{\plotpanel}[3]
           {\vbox{\hsize=#2\epsfig{file=#1,width=#2}
                  \ifnum#3=1\makebox[\hsize]{\thepanel}\fi}}

\slugcomment{To appear in the Astrophysical Journal}

\begin{document}
\fontsize{10}{10.6}\selectfont
\title{Centimeter Imaging of the R Coronae Australis Region}
\author{\sc Minho Choi\altaffilmark{1}, Kenji Hamaguchi\altaffilmark{2,3},
            Jeong-Eun Lee\altaffilmark{4},
            and Ken'ichi Tatematsu\altaffilmark{5}}
\altaffiltext{1}{International Center for Astrophysics,
                 Korea Astronomy and Space Science Institute,
                 Hwaam 61-1, Yuseong, Daejeon 305-348, South Korea;
                 minho@kasi.re.kr.}
\altaffiltext{2}{CRESST and X-ray Astrophysics Laboratory NASA/GSFC,
                 Greenbelt, MD 20771.}
\altaffiltext{3}{Department of Physics, University of Maryland,
                 Baltimore County, 1000 Hilltop Circle, Baltimore, MD 21250.}
\altaffiltext{4}{Department of Astronomy and Space Science,
                 Astrophysical Research Center
                 for the Structure and Evolution of the Cosmos,
                 Sejong University, Gunja 98, Gwangjin,
                 Seoul 143-747, South Korea.}
\altaffiltext{5}{National Astronomical Observatory of Japan,
                 2-21-1 Osawa, Mitaka, Tokyo 181-8588, Japan.}
\setcounter{footnote}{5}

\begin{abstract}
\fontsize{10}{10.6}\selectfont
The R CrA region was observed in the 3.5 and 6.2 cm continuum
with high angular resolutions (0.6--1.7 arcseconds)
using the Very Large Array.
Archival data sets were also analyzed for comparison,
which provided angular resolutions up to 0.3 arcseconds.
A cluster of young stellar objects was detected,
and a rich array of star forming activities was revealed.
IRS 7A showed an enhanced outflow activity recently.
The main peak of IRS 7A positionally coincides with an X-ray source,
which suggests that the X-ray emission
is directly related to the central protostar.
The Class 0 source SMA 2 is
associated with a double radio source, B 9a and 9b,
and seems to be driving two outflows.
The B 9 complex is probably a multiple-protostar system.
Both B 9a and 9b are nonthermal radio sources
with negative spectral indices.
IRS 7B is a compact radio source surrounded by an extended structure.
The compact source corresponds to the Class 0/I source SMA 1,
and it is also closely associated with an X-ray source,
suggesting that magnetic activities start early
in the protostellar stage of evolution.
The extended structure of IRS 7B may be a bipolar outflow.
IRS 5 was resolved into two sources with a separation of 0.9 arcseconds.
Both IRS 5a and 5b display radio flares and X-ray emission,
suggesting that energetic magnetic processes are active in both members.
The month-scale active phase of IRS 5b implies
that the flare activity must involve large-scale magnetic fields.
During the strong flare event of IRS 5b in 1998,
IRS 5a also showed an enhanced level of radio emission.
This concurrent activity suggests
that IRS 5 may be an interacting young binary system,
but the interaction mechanism is unknown.
Alternatively, what was seen in the radio images
could be a circumbinary halo.
The variable radio source B 5 was found to be a nonthermal source at times,
and the size of the 6.2 cm source is about 1 arcsecond,
suggesting that B 5 is a Galactic object.
A radio outburst of IRS 6 was detected once,
and the radio/X-ray source was identified as IRS 6a.
The other member of the IRS 6 system, IRS 6b, was undetected in X-rays,
suggesting that only IRS 6a has detectable magnetic activities.
Properties of other radio sources, IRS 1, IRS 2, and R CrA, are discussed,
and the radio detections of T CrA and WMB 55 are reported.
The proper motion of R CrA was marginally detected.
Also presented is the classification of infrared sources
in the R CrA region based on an infrared color-color diagram.
\end{abstract}


\section{INTRODUCTION}

The Corona Australis molecular cloud is
one of the nearest star-forming regions,
and the R CrA region is
the most active site of star formation in this cloud
(Harju et al. 1993; Chini et al. 2003).
The R CrA cloud core contains a cluster of young stellar objects
known as the ``Coronet'' cluster
(Taylor \& Storey 1984; Wilking et al. 1992, 1997).
Deeply embedded protostars and preprotostellar cores were detected
in the (sub)millimeter continuum
(Henning et al. 1994; Choi \& Tatematsu 2004;
Nutter et al. 2005; Groppi et al. 2004, 2007).
Star forming activities,
such as infall, accretion, molecular outflows,
and Herbig-Haro objects,
were revealed by millimeter spectroscopic observations and optical imaging
(Strom et al. 1974; Hartigan \& Graham 1987; Anderson et al. 1997;
Groppi et al. 2004, 2007; Nisini et al. 2005; Wang et al. 2004).
Centimeter continuum imaging with interferometers
has been especially effective
in probing the nature of the deeply embedded objects
and determining their precise positions
(Brown 1987; Suters et al. 1996; Feigelson et al. 1998;
Forbrich et al. 2006; Miettinen et al. 2008).
High-energy processes around some of the young stellar objects
were discovered in X-ray observations (Koyama et al. 1996;
Hamaguchi et al. 2005, 2008; Forbrich et al. 2006, 2007;
Forbrich \& Preibisch 2007).
The distance to the CrA molecular cloud is $\sim$170 pc
(Knude \& H{\o}g 1998).

Among the near-IR sources discovered by Taylor \& Storey (1984)
in the R CrA cloud core,
IRS 7 is the most deeply embedded and probably the youngest object.
In the centimeter continuum imaging by Brown (1987),
IRS 7 was resolved into three radio sources: IRS 7A, IRS 7B, and B 9.
Single-dish observations of submillimeter continuum emission
suggested that they are deeply embedded objects (Nutter et al. 2005).
Several radio continuum sources were detected around IRS 7A
(Choi \& Tatematsu 2004; Forbrich et al. 2006),
suggesting that this small region contains a multiple-protostar system.
The IRS 7A region is interesting
because it appears to contain objects with inhomogeneous spectral classes
or in different stages of evolution (Choi \& Tatematsu 2004).
While IRS 7A appears to be a Class I protostar
as it was detected in near-IR,
the other members of the multiple system (CT 3 and B 9)
were suggested to be probable Class 0 protostars.
Recent interferometric observations in the millimeter continuum confirmed
that B 9 (SMA 2) is a Class 0 source
while IRS 7A and CT 3 were undetected (Groppi et al. 2007).

\input{tab1.tex}

IRS 7B attracted attention
because it is a rare example of Class 0 protostars showing X-ray emission
(Feigelson \& Montmerle 1999; Hamaguchi et al. 2005).
Recent interferometric observations in the millimeter continuum suggested
that IRS 7B (SMA 1) may be a Class 0/I transitional object
(Groppi et al. 2007).
Hamaguchi et al. (2005) demonstrated
that the X-ray emission from IRS 7B comes from two different components:
one hot and variable, and the other cool and constant.
The cool component was interpreted as originating from collisional shock
of the protostellar jet with circumstellar gas,
and such X-ray emission has been found around several Class 0/I protostars,
for example, HH 2, L1551 IRS 5, and OMC 3 MMS 2
(Pravdo et al. 2001; Favata et al. 2002; Bally et al. 2003;
Tsujimoto et al. 2004).
The hot variable X-ray emission
had not been detected from other Class 0 protostars.
This component would be related to the reconnection of magnetic fields
around the protostar during mass accretion (e.g., Montmerle et al. 2000).
Spatially separating the jet collision region from the central protostar
with a high resolution imaging
is therefore important in understanding the structure
of the protostellar system.

IRS 5 is an interesting young binary system.
It was spatially resolved in the near-IR and X-ray images
(Nisini et al. 2005; Hamaguchi et al. 2008).
Both members of the system show quiescent X-ray emission,
but flare activities of them were remarkably different:
only IRS 5a frequently showed solar type X-ray flares
(Hamaguchi et al. 2008).
IRS 5 also displays radio outbursts
and emits circularly polarized radio emission
(Suters et al. 1996; Feigelson et al. 1998;
Forbrich et al. 2006; Miettinen et al. 2008).
However, the radio properties of each binary member
need to be characterized separately to understand the radio activities.

The R CrA region also contains other radio sources
displaying interesting star-formation activities of their own.
IRS 1 and IRS 2 are Class I protostars and variable X-ray sources
(Forbrich et al. 2006).
IRS 1 is the driving source of the HH 100 outflow (Strom et al. 1974).
WMB 55 is a Class I source without an X-ray detection
(Wilking et al. 1997; Nutter et al. 2005).
IRS 6 is probably an X-ray emitting T Tauri star
driving a giant Herbig-Haro flow
(Wang et al. 2004; Forbrich et al. 2006).
R CrA and T CrA are Herbig Ae/Be stars,
and each of them may be a binary system (Takami et al. 2003).
B 5 is a radio source of unknown nature, and its flux varies rapidly
(Brown 1987; Suters et al. 1996; Feigelson et al. 1998).

To obtain high-quality images of the young stellar objects
despite the low declination,
we observed the R CrA region with the Very Large Array (VLA)
using a fast-switching phase calibration technique.
In this paper, we present
our centimeter continuum observations of the R CrA region.
We describe our radio continuum observations in \S~2
and archival data in \S~3.
In \S~4 we briefly report the results of the continuum imaging.
In \S~5 we discuss the star forming activities
of the radio sources in the R CrA region.
A summary is given in \S~6.

\section{OBSERVATIONS}

The R CrA IRS 7 region was observed
using VLA of the National Radio Astronomy Observatory (NRAO)
in the standard $X$-band continuum mode (8.5 GHz or $\lambda$ = 3.5 cm)
and in the standard $C$-band continuum mode
(4.9 GHz or $\lambda$ = 6.2 cm).
Twenty-three antennas were used in the BnA-array configuration
on 2005 February 3 for the 3.5 cm observations
and on 2005 February 12 for the 6.2 cm observations
(Tr 13/14 in Table 1).
The phase-tracking center was
$\alpha_{2000}$ = 19$^{\rm h}$01$^{\rm m}$55.33s,
$\delta_{2000}$ = --36\arcdeg57$'$21.66$''$.
The phase was determined
by observing the nearby quasar 1924--292 (QSO B1921--293).
The target-calibrator cycle was 4 and 5 minutes
for the 3.5 and 6.2 cm observations, respectively.
In the 3.5 cm observations, the flux was calibrated
by setting the flux density of 1924--292 to 13.7 Jy,
which is the flux density measured within 16 days of our observations
(VLA Calibrator Flux Density Database%
\footnote{See http://aips2.nrao.edu/vla/calflux.html.}).
To check the flux scale,
the quasar 1331+305 (4C 30.26) was observed in the same track.
Comparison of the amplitude gave a flux density of 5.30 Jy for 1331+305,
which agrees with the value in the VLA Calibrator Flux Density Database
to within 3\%.
In the 6.2 cm observations, the flux was calibrated
by setting the flux density of 1331+305 to 8.25 Jy,
which is the flux density measured within 2 days of our observations
according to the VLA Calibrator Flux Density Database.
Comparison of the amplitude gave a flux density of 13.52 Jy for 1924--292.
Because of the short cycle time,
the quality of phase calibration was good.
We tried a self-calibration,
but it did not make a noticeable difference
in the quality of resulting images.
Therefore, the Tr 13/14 images presented in this paper
were made without self-calibration.
Maps were made using a CLEAN algorithm.

\section{ARCHIVAL DATA}

In addition to our data, we also analyzed several VLA data sets
retrieved from the NRAO Data Archive System.
We searched for 3.5/6.2 cm continuum observations of the R CrA region
made in the A-, BnA-, or B-array configuration
and found four observing projects in recent years.
Including our observations,
we present the results of 14 observing runs.
A summary of the observing runs is given in Table 1.
The archival VLA data were obtained from observations
with a typical target-calibrator cycle time of 10--15 minutes.
To improve the phase calibration
we used the phase-only self-calibration technique
(see Forbrich et al. 2006).
The Tr 1--12 images presented in this paper were made
with the self-calibration applied.

\subsection{Project AB 807}

The NRAO observing project AB 807 includes a single observing run (Tr 1).
The data set contains data from the observations of the R CrA IRS 7 region
in the standard $X$-band continuum mode
and in the standard $C$-band continuum mode in a single track.
Twenty-five antennas were used in the A-array configuration.
The phase-tracking center was
$\alpha_{1950}$ = 18$^{\rm h}$58$^{\rm m}$33.0s,
$\delta_{1950}$ = --37\arcdeg01$'$41.0$''$.
The phase calibrators were the quasars QSO B1921--293 and QSO B1933--400,
and the flux calibrator was the quasar 3C48.
The bootstraped flux densities were
17.4 Jy for QSO B1921--293 and 0.96 Jy for QSO B1933--400 in the $X$-band
and 
16.2 Jy for QSO B1921--293 and 0.98 Jy for QSO B1933--400 in the $C$-band.

\subsection{Project AC 481}

The NRAO observing project AC 481 includes two observing runs (Tr 2 and 3).
Details of the observations and results
were presented by Feigelson et al. (1998).
The flux densities of sources in the images produced by us
agree well with the values reported by Feigelson et al. (1998).
However, the positions of compact sources have an offset
compared with those found in the images of the other data sets.
The amount of the position offset is
0.1$''$ to the east and 0.4$''$ to the north.
A probable cause of this offset is
that the phase calibrator used in the project AC 481 was 1744--312.
In the other projects
the phase calibrator was 1924--292, as in our observations.
The position accuracy of 1744--312 is worse than that of 1924--292:
1744-312 is a code-C calibrator while 1924--292 is a code-A calibrator.
The angular distance from the target source is also larger:
17.5\arcdeg\ for 1744--312 and 9.2\arcdeg\ for 1924--292.
Therefore, we conclude that the astrometry of the AC 481 images
is less reliable than that of the other images.
In all the figures presented in this paper,
a position correction was applied to the AC 481 images.

\subsection{Projects AK 469 and AM 596}

The NRAO observing projects AK 469 and AM 596
include nine observing runs (Tr 4--12).
Details of the observations and results
were presented by Forbrich et al. (2006).
The flux densities and positions of compact sources
in the images produced by us
agree well with the values reported by Forbrich et al. (2006).

\subsection{Spitzer Data}

In addition to the radio data,
we analyzed archival infrared data
to classify the young stellar objects in the R CrA region.
From the {\it Spitzer} data archive,
GTO observations for the R CrA region
(AOR keys: IRAC 3650816 and MIPS 3664640) were retrieved.
The aperture photometry was performed with mosaic post-BCD images,
which were processed by the SSC pipeline
(Ver. S14 and S16 for IRAC and MIPS, respectively),
at all IRAC bands (3.6, 4.5, 5.8, and 9.0 $\mu$m)
and MIPS 24 and 70 $\mu$m using the SSC Mopex package.
The short-exposure high-dynamic range observations
were used for the IRAC photometry.

\section{RESULTS}

\input{tab2.tex}

Table 2 lists all the 16 sources detected in the radio maps.
Figure 1$a$ shows the average map of the R CrA region
in the 3.5 cm continuum from all the data sets,
and Figure 1$b$ shows the maps of the IRS 7 region from our observations.
All the images presented in this paper
are corrected for the primary beam response,
and the lowest contour level roughly corresponds to the 3$\sigma$ value,
where $\sigma$ is the rms of noise.
Objects associated with the centimeter continuum sources
are marked in the figures,
and the size of markers roughly corresponds
to the position uncertainty of the objects.
Tables 2 and 3 list the parameters of detected sources,
and Figure 2 shows the flux variability of bright sources.
We briefly summarize
the overall properties of detected sources in this section,
and detailed discussion on individual sources will be given in \S~5.

The IRS 7 region shows a complex of several compact sources
and some extended emission structures.
Since there are several published papers
reporting the centimeter continuum sources in this region
(many of them have their own list of sources),
simply using source numbers can be confusing.
In this paper, the source names listed in Table 3 are used.
We tried to list the most commonly used names.
In some cases, for example, a name may be a combination of
an acronym (names of authors who identified the source first)
and a sequence (source number) used by those authors.

\subsection{Source Structure}

IRS 7A is the strongest centimeter-continuum source in this region
and is composed of a compact source and an extended emission structure.
B 9 is associated with the 7 mm source CT 2 and the 1.1 mm source SMA 2
and is probably the most deeply embedded source in this region.
B 9 was resolved into two compact sources.
FPM 10 and FPM 13 are extended emission structures,
and their shapes and peak positions
change depending on the $uv$ coverage and the epoch of observations.

IRS 7B is also composed of a compact source
surrounded by an extended emission structure.
The compact source is associated with the 1.1 mm source SMA 1.
The extended structure clearly has a bipolar morphology
elongated in the northeast-southwest direction.
The northeastern lobe corresponds to FPM 15 (Forbrich et al. 2006)
and shows a clumpy structure.

IRS 5 was resolved into two compact sources,
and they displayed interesting flux variability.
B 5 is slightly extended in the northeast-southwest direction,
and the deconvolved size along the elongation is $\sim$1$''$.
Most of the other detected sources in the R CrA field
are point-like objects.

\subsection{Spectral Energy Distribution}

Table 3 lists $\alpha$, the spectral index in the centimeter band.
For most sources, the spectral index
is consistent with the value expected for the free-free emission
from collimated thermal radio jets or coronal winds
($-0.1 \leq \alpha \lesssim 1$; Reynolds 1986; Anglada et al. 1998).
However, there are a few exceptions.
The spectral index of B 9 is negative
in both Tr 1 and Tr 13/14 estimates,
suggesting that the centimeter emission
is from a nonthermal process at least partially.
The spectral index of IRS 7B needs some discussion (see \S~5.2).
The spectral index of B 5 varies wildly,
which is probably due to a combined effect
of nonthermal emission processes and flare activities (see \S~5.4).

It should be noted that,
even though the spectral index of IRS 5 is consistent with free-free emission,
its circularly polarized flux certainly tells
that it has a nonthermal emission component.
There are other examples of nonthermal emitter
with a slightly positive spectral index,
such as T Tau Sb (Loinard et al. 2007).
These sources suggest that the spectral index alone
is not a very reliable indicator of emission process.
On the other hand, in radio surveys of star forming clouds,
sources displaying negative spectral indices
are usually considered background nonthermal sources
(e.g., Rodr{\'\i}guez et al. 1999),
but the existence of young stellar objects such as B 9
tells that the spectral index alone is not a good criterion
for distinguishing young stellar objects from extragalactic objects.

\subsection{Long-Term Variability}

Figure 2 shows long-term (8--13 years) variability
of the 3.5 cm flux density.
Considering the month-scale variability,
most sources do not show clear trends.
IRS 7A seems to show a steady increase of flux density.
More observations are needed to confirm this trend.
If real, such long-term variability
is likely due to changes in outflow activity
rather than circumstellar magnetic activities.

\input{fig1.tex} \input{tab3.tex} \input{fig2.tex}

\subsection{Proper Motion}

Proper motions of bright compact sources were measured
by comparing the uniform-weight 3.5 cm maps of Tr 1 and Tr 13.
Most sources did not show a detectable proper motion.
R CrA is the only source
showing a marginally detectable proper motion:
0.23$''$ $\pm$ 0.11$''$ toward the southeast over the 8.1 yr time-baseline,
or (20, --21) mas yr$^{-1}$,
which is consistent with the proper motion of the R CrA association,
(5.5, --27.0) mas yr$^{-1}$ (Neuh{\"a}user et al. 2000).

\subsection{Polarimetry}

Circularly polarized 3.5 cm emission was detected toward three sources
while none of the radio sources
showed detectable polarization in the 6.2 cm band.
IRS 5b showed detectable polarization in most observing runs,
but IRS 5a was never detected in the Stokes-$V$ flux maps.
Polarized emission from IRS 7A and B 5 was detected once each.
Details of the polarimetry results will be presented separately
in Paper II (Choi et al. 2008).

\subsection{Infrared Photometry}

\input{tab4.tex}

\input{fig3.tex}

Since flux measurements and images based on {\it Spitzer} data
are reported previously for many of the sources in the R CrA region
(Groppi et al. 2007; Forbrich \& Preibisch 2007),
we report the photometry of IRS 2 and IRS 6 only (Table 4).
The IRAC color-color diagram for the infrared sources in the R CrA region
is shown in Figure 3.

For IRS 2 and IRS 6,
the bolometric temperature and luminosity
were calculated using the {\it Spitzer} data
along with the Two Micron All Sky Survey (2MASS) data.
Although IRS 2 has a greater bolometric luminosity
than IRS 6 by a factor of 25,
they have a similar bolometric temperature of $\sim$450 K,
indicative of Class I sources (Chen et al. 1995).
According to the [3.6]--[4.5] versus [5.8]--[8.0] color-color diagram
(Allen et al. 2004; Lee et al. 2006),
both IRS 2 and IRS 6 are also classified
as embedded Class 0/I sources (Fig. 3).

\section{DISCUSSION}

\subsection{IRS 7A Region}

\input{fig4.tex}

\input{fig5.tex}

\input{fig6.tex}

The IRS 7A region is a very complicated site of star formation.
There are at least two young stellar objects in this small region,
IRS 7A and B 9 (SMA 2),
and each of them have subcomponents.
All these sources seem to be embedded
in the dense core SMM 1C%
\footnote{In this paper,
we use the submillimeter source numbers given by Nutter et al. (2005).
Choi \& Tatematsu (2004) referred to SMM 1A, 1B, and 1C
as vdA 4, 5, and 3, respectively,
following the source numbers given by van den Ancker (1999).
There is a difference in astrometry
between Nutter et al. (2005) and Groppi et al. (2007),
and we use the positions given by Groppi et al., if possible.}
(Nutter et al. 2005; Groppi et al. 2007).
Figures 4--6 show the centimeter continuum maps of this region.
We will try to dissect this region
by comparing the radio maps with the images in other wavelengths.

\subsubsection{IRS 7A}

The Tr 1 map of Figure 6 has the highest resolution
and shows the accurate position of the compact component of IRS 7A.
In addition to the compact component,
the other maps of Figure 6 as well as the average map (Fig. 4)
show the extended structure
that seems to be elongated roughly in the north-south direction.
Judging from the spectral index and the morphology,
the centimeter emission of IRS 7A most likely traces a thermal radio jet.
The outflow activity may have become enhanced in recent years
as the 3.5 cm flux density in 2005 (Tr 13) is significantly higher
than those in the 1996--1998 period (Fig. 2).
The 3.5 cm map of Tr 13 (Fig. 6) also shows
a newly appearing subcomponent at 0.85$''$ southeast of the main peak,
which may be a bright knot of recently ejected material.
Since this subcomponent is not present in the images of earlier epoch,
its proper motion is at least 130 mas yr$^{-1}$
($\sim$100 km s$^{-1}$ at 170 pc).

The peak position of the 3.5 cm maps
has not changed over the 8 years of monitoring
and may indicate the location of the central object.
As shown in Figure 6, the 3.5 cm source position agrees well
with the positions of the sources detected in other wavelengths:
the mid-IR source IRAC 5, the 7 mm source CT 4,
and a {\it Chandra} X-ray source
(Groppi et al. 2007; Choi \& Tatematsu 2004; Hamaguchi et al. 2005).
Despite all these detections over a wide range of wavelength,
the classification (hence the evolutionary status) of IRS 7A
is frustratingly uncertain because of the complicated environment.
IRS 7A is certainly not a T Tauri star.
The nondetection at 1.1 mm (Groppi et al. 2007)
seems to rule out the possibility of IRS 7A being a Class 0 source,
and Hamaguchi et al. (2005) suggested
that IRS 7A may be a Class I source seen at a large inclination angle
based on the near-IR magnitude
and the column density measurement from X-ray observations.
The IRAC color-color diagram places IRS 7A near IRS 7B
and above the other Class I sources (Fig. 3),
which suggests that IRS 7A may be
either a Class 0/I transitional object
or a Class I object in the younger part of the class.

Hamaguchi et al. (2005) could reproduce
the {\it XMM-Newton} spectrum of the X-ray source X$_{\rm W}$
using a single-temperature plasma model,
which could also reproduce the {\it Chandra} data.
Comparing the 3.5 cm maps with the {\it Chandra} X-ray positions (Fig. 6),
the X-ray source seems to be associated with the main radio peak itself
but not with the subcomponent on the southeast side
or the more extended component on the south side of the main peak.
(The {\it Chandra} observations were made in 2000 October and 2003 June.)
This geometry suggests that the X-ray source
may be closely related with the central star.
The X-ray emission may come from either
the protostellar magnetosphere or near the base of outflow.
The detection of circularly polarized radio emission (Paper II)
strongly suggests the existence of an active magnetosphere.

\subsubsection{B 9 Complex}

The centimeter source B 9 was resolved into two sources, B 9a and 9b,
with a separation of 0.5$''$ (Fig. 6).
B 9a is slightly elongated in the east-west direction
with a deconvolved size of $\sim$0.5$''$,
and B 9b is a point-like source.
Both of them show negative spectral indices (--0.9 to --0.2, see Table 3),
suggesting that the centimeter emission
comes from a nonthermal process at least partially.
In contrast, Miettinen et al. (2008) reported positive (0.4) spectral index,
which suggests that either the nonthermal component is highly variable
or their flux measurement included
a significant contribution from the extended thermal emission source.
The proper motion of B 9a/b is uncertain
because they were not clearly separated in the Tr 13 image.

The centimeter source B 9 and the 7 mm source CT 2
are located close to each other,
but not exactly at the same position.
It is unlikely that this position difference (1.2$''$)
is caused by an instrumental uncertainty
because the position agreement of IRS 7A and CT 4 is good.
The peak position of the 1.1 mm source SMA 2 is 0.4$''$ south of B 9a,
but B 9b and CT 2 are also located
within the beam size of the 1.1 mm map of Groppi et al. (2007).
We will use the name `B 9 complex'
to refer to all these compact sources collectively.
The spectral energy distribution (SED) of the B 9 complex
(see Fig. 8 of Groppi et al. 2007) suggests
that it may be harboring a Class 0 source,
but it is not clear what is the true nature of the B 9 complex.
Considering the complexity of the centimeter-millimeter maps
and the existence of two outflows (see below),
we suggest that the B 9 complex may be a multiple-protostar system.
A subarcsecond-resolution map of dust emission
is needed to locate the central stellar object(s)
and understand the situation better.

Radio sources with significantly negative spectral index
are rare in star forming regions but not unique.
Examples include outflow lobes of Serpens SMM 1 and IRAS 16547--4247
(Rodr{\'\i}guez et al. 1989; Garay et al. 2003).
The suggested radiation mechanism is nonthermal synchrotron emission,
but the existence of highly relativistic electrons is not easy to explain.
It is possible to accelerate a small fraction of electrons
in the shocked region of outflow (e.g., Crusius-W{\"a}tzel 1990),
but the density of the medium should be low enough
to avoid being dominated by thermal radiation (Henriksen et al. 1991).
It is also possible to produce energetic electrons in active magnetospheres,
as observed around more evolved young stellar objects
(Massi et al. 2006, 2008; Loinard et al. 2007).
Other types of nonthermal radiation (such as gyrosynchrotron radiation,
electron cyclotron maser, and plasma maser)
cannot be ruled out (G{\"u}del 2002),
but they are less likely
because circular polarization was not detected toward B 9 (Paper II).
It remains to be seen if B 9a/b
are indeed shocked regions of outflowing gas,
and measurements of proper motion and linear polarization
will be helpful.

\subsubsection{FPM 10 and FPM 13}

FPM 10 and 13 are extended sources,
and their shape and peak position
change depending on the epoch of observations.
Their centimeter spectra are flat,
suggesting optically thin free-free emission.
They are highly elongated (Figs. 1, 4, and 5),
and the major axes point to the B 9 complex.
Therefore, they are most likely ionized gas of outflows
driven by young stellar object(s) in the B 9 complex.
However, they cannot be a single bipolar outflow system
because their directions with respect to B 9
are nearly perpendicular to each other.
Forbrich \& Preibisch (2007) found diffuse X-ray emission
in a region just north of FPM 13,
but the relation between the extended X-ray structure and the B 9 complex
is not clear.

FPM 10 is located about 3$''$ west of the B 9 complex
and elongated in the east-west direction.
The size along the major axis is $\sim$3$''$.
The location and the flow direction of FPM 10 match well
with the redshifted lobe of the bipolar outflow
mapped in several molecular lines
including HCO$^+$ $J$ = 1$\rightarrow$ 0 and CO $J$ = 3 $\rightarrow$ 2
(Anderson et al. 1997; Groppi et al. 2004, 2007).
Therefore, FPM 10 and the molecular bipolar outflow
seem to belong to a single outflow system.

FPM 13 is located about 4$''$ north of the B 9 complex
and elongated in the north-south direction.
The size along the major axis is $\sim$3$''$.
There is no known molecular outflow that can be associated with FPM 13,
but this is probably because the molecular line maps
do not have a sufficiently high angular resolution.
Because of the flow direction,
the driving source of FPM 13 may be an object in the B 9 complex
but cannot be the same object that is driving FPM 10.
It should also be noted
that we cannot completely rule out the possibility
that FPM 13 may be driven by IRS 7A.

\subsubsection{CT 3}

CT 3 was nearly as bright as CT 2 in the 7 mm map
(Choi \& Tatematsu 2004)
but not clearly detected in the centimeter maps.
It is located in the ``bridge'' of radio emission structure
between IRS 7A and B 9 (Fig. 4).
This ``bridge'' may be tracing an outflow
either from IRS 7A or from the B 9 complex
(or a superposition of outflows from both).
A marginally detectable amount of 3.5 cm emission
can be seen in the Tr 13 map (Fig. 1$b$ {\it left panel}).
Choi \& Tatematsu (2004) suggested
that CT 3 may be a deeply embedded young stellar object,
but the nondetection at 1.1 mm (Groppi et al. 2007)
seems to rule out this possibility.
A possible explanation is
that CT 3 may be a knot of outflowing gas
that was temporarily brightened at the time of the 7 mm observations.

\subsection{IRS 7B}

The IRS 7B region is less complicated than the IRS 7A region
but still needs a careful analysis to understand the structure.
Figures 7--9 show the centimeter continuum maps of the IRS 7B region.
There is a single compact source surrounded by an extended structure
elongated in the northeast-southwest direction.
The spectral index was highly negative (--1.4) during Tr 1
but only mildly negative (--0.2) during Tr 13/14 (Table 3).
This difference is due to the low 3.5 cm flux during Tr 1,
which probably means that the extended structure
was resolved out in the A-array observations.
If only the compact source is considered,
the spectral index is $\sim$0.1 during Tr 1 and $\sim$0.7 during Tr 13/14.
Therefore, the centimeter flux from the central compact source of IRS 7B
may be mostly from free-free emission.

\input{fig7.tex}

\input{fig8.tex}

\input{fig9.tex}

The peak position of the 3.5 cm maps
did not show a detectable proper motion over the 8 years of monitoring
and may indicate the location of the central object.
As shown in Figure 9,
the 3.5 cm peak position agrees with the positions
of the mid-IR source IRAC 4, the 1.1 mm source SMA 1,
and the {\it Chandra} X-ray source
(Groppi et al. 2007; Hamaguchi et al. 2005).
Groppi et al. (2007) suggested that IRS 7B is a Class 0/I source
based on the SED analysis.
Therefore, IRS 7B may be a protostar embedded in the dense core SMM 1B.

The extended component of IRS 7B shown in Figure 7
has a length of $\sim$9$''$ and a position angle of $\sim$20\arcdeg.
The bipolar morphology suggests
that the extended component traces a bipolar outflow driven by IRS 7B.
The northeastern lobe (FPM 15) is brighter than the opposite lobe.
The outflow in our maps
is probably the base of the large-scale ($\sim$2$'$) bipolar jet
in the 6 cm maps of Brown (1987) and Miettinen et al. (2008).
Large-scale maps of molecular lines from single-dish observations
did not show any sign of molecular outflow parallel to the IRS 7B outflow.
Again, this nondetection is
probably caused by a confusion with the east-west outflow of B 9.
The interferometric map in the HCO$^+$ line (Fig. 6 of Groppi et al. 2007)
seems to show a molecular outflow around SMA 1:
blueshifted emission on the south of SMA 1
and redshifted emission on the northeast of SMA 1.
The size and direction of the HCO$^+$ emission structure
seem to agree with the outflow seen in our centimeter continuum maps.
However, note that Groppi et al. (2007) interpreted
the HCO$^+$ emission structure as a part of a large spherical shell.

Similarly to IRS 7A, the {\it Chandra} X-ray source
is associated with the main peak of the centimeter source, IRS 7B,
but not with the secondary peaks in the outflow lobe such as FPM 15.
This geometry suggests that the X-ray source
is associated with the central protostar.
Hamaguchi et al. (2005) found
that the {\it XMM-Newton} spectra of X$_{\rm E}$
could be reproduced better using a two-temperature model.
They concluded that the hard X-ray component
comes from a mass accretion spot on the protostar
while the soft component probably comes from shock-heated plasma
related to a jet/outflow.
The high plasma temperature and the large X-ray luminosity
require a large amount of gas at a very high outflow velocity
(Hamaguchi et al. 2005; Forbrich et al. 2006).
A possible model reconciling the radio and the X-ray observations is
that the hard and time-variable X-ray component
may be emitted by circumstellar magnetic activity
and that the soft and constant X-ray component and the centimeter emission
may come from plasma
heated by the collision of a steady jet with ambient gas
(Hamaguchi et al. 2005; Forbrich \& Preibisch 2007).
Since the X-ray spectra from {\it Chandra} observations
were dominated by the soft component
and the spatial resolution of {\it XMM-Newton} is limited ($\sim$6$''$),
the exact position of the hard component is unknown.
To understand the IRAS 7B system clearly,
it is necessary to observe the X-ray source with a high resolution
during an active phase of the hard component.

\subsubsection{CHLT 13}

CHLT 13 is a 6.2 cm continuum source detected during Tr 14 (Fig. 1$b$).
Since it was not detected in the 6.2 cm map of Tr 1
and never detected in the 3.5 cm continuum,
it must be a transient source.
Its location is near the dense core SMM 1A.
Nutter et al. (2005) suggested
that the submillimeter source SMM 1A may be a prestellar object
because it was not detected in other wavelengths.
On the other hand, the 6 cm maps of Miettinen et al. (2008) show
extended clumpy structure around this region,
but none of their emission peaks coincides with CHLT 13 exactly.
On the nature of CHLT 13, we suggest two possibilities.
First, CHLT 13 may indicate a star forming activity in SMM 1A.
Second, CHLT 13 may be a temporarily brightened knot of the IRS 7B outflow.
More speculatively, the southwestern outflow of IRS 7B
may be colliding with a dense clump in SMM 1A
and producing transient radio emission at the shock front.

\subsection{IRS 5}

\input{fig10.tex}

\input{fig11.tex}

IRS 5 is a very interesting binary system.
Figures 10--12 show the centimeter continuum maps of the IRS 5 region.
IRS 5 was clearly resolved into two sources
when the resolution is high enough
(see the maps of Tr 1, Tr 2, and Tr 13 in Fig. 12).
Even when the resolution is not high enough,
the 3.5 cm maps usually show an extended structure
that can be best interpreted as a combination of two point-like sources
(see the maps of Tr 4 and Tr 6--12 in Fig. 12).
In addition, comparison with the Stokes-$V$ maps (Paper II)
clearly shows that the southwestern source
was detected in the Stokes-$I$ maps.
Since the 3.5 cm sources are
similar to the near-IR sources (Chen \& Graham 1993; Nisini et al. 2005)
in separation and position angle,
we identify the southwestern and the northeastern sources
as IRS 5a and 5b, respectively.
The Stokes-$V$ maps show a point-like source at the position of IRS 5b
while IRS 5a was undetected in the Stokes-$V$ flux,
and there is a good correlation
between the Stokes-$V$ and Stokes-$I$ fluxes of IRS 5b
(see the discussion in Paper II).

IRS 5a is brighter in the near-IR images of Nisini et al. (2005),
but IRS 5b is usually brighter in radio.
The separation between them in the 3.5 cm Tr 1 image
is 0.9$''$ (160 AU at 170 pc).
They were not clearly separated in the 6.2 cm maps,
but the Tr 1 map (Fig. 11 {\it left panel}) shows
that both IRS 5a and 5b have detectable 6.2 cm fluxes.
Since circularly polarized flux was detected from IRS 5,
the centimeter flux clearly has a nonthermal component,
and we will discuss this issue in Paper II.
Another Class I source in this region, IRS 5N (FPM 5),
was not clearly detected
though Forbrich et al. (2006) reported
that it was detected when several data sets were combined.

\input{fig12.tex}

Table 5 lists the 3.5 cm flux densities of each binary component.
For the maps of Tr 4--12,
the flux was measured by fitting the image
with a sum of two point-like sources
(see the end note of Table 5 for details).
The uncertainty of total flux from this fitting method
was estimated by performing simple Monte Carlo simulations.
Since we assumed that both sources are point-like,
the uncertainty can be larger if this assumption is not valid.
The light curves of IRS 5a and 5b are shown in Figure 13.
The flux uncertainty shown in Figure 13
is the statistical uncertainty listed in Table 5,
but one should also consider the uncertainty in the absolute flux scale,
which was not explicitly measured but usually about 1--2\%
(VLA Calibrator Manual%
\footnote{See http://www.vla.nrao.edu/astro/calib/manual.}).
The radio light curves clearly show that both sources are highly variable.
IRS 5b is usually brighter and more active than IRS 5a,
but IRS 5a outshone its companion at least once (Tr 4)
during the monitoring.
This behavior clearly tells
that the flux variability of each source should be analyzed separately,
but previous authors considered the flux of the IRS 5 system as a whole,
which may have caused some confusion.

The radio positions of IRS 5a/b agree well
with the position of X-ray sources (Hamaguchi et al. 2008).
The separation between the X-ray sources is $\sim$0.8$''$,
which is in good agreement with the radio value.
There seems to be a difference
in the relative degree of flare activity between the two sources.
IRS 5b is more active in radio (Fig. 13),
but IRS 5a is more active in X-rays (Hamaguchi et al. 2008).
To understand this difference,
it is necessary to monitor the IRS 5 system
in both radio and X-rays simultaneously.
Forbrich et al. (2007) attempted such a study,
but IRS 5 was not active in radio during their observations
and their angular resolution was not high enough to resolve the system.

The position and relative brightness of the near-IR counterparts
need some consideration.
Nisini et al. (2005) reported
that the separation between the two near-IR sources is $\sim$0.6$''$,
but their Figure 1 shows that the separation is actually $\sim$0.9$''$,
in agreement with the radio/X-ray value.
The absolute position of the near-IR sources
in the image of Nisini et al. (2005)
has an offset of $\sim$1$''$ to the southeast
relative to the radio/X-ray position.
We presume that this offset was caused
by an instrumental uncertainty of the near-IR observations.
It is interesting to note
that the 2MASS position agrees with the radio position of IRS 5b
while the mid-IR source IRAC 10 coincides with IRS 5a (Fig. 10).
This difference between the 2MASS and the IRAC positions
is not an astrometric problem
because the positions of IRS 5N show a good agreement.
Therefore, we suppose that the primary object in the near-IR brightness
can be either one depending on the epoch of observations.
That is, either one or both of them may be variable in near-IR.
Alternatively, they may have different intrinsic colors.
The relation between the variability in near-IR and in radio/X-rays
has not been well understood so far (Forbrich et al. 2007).

\input{tab5.tex}

\input{fig13.tex}

IRS 5a/b (and IRS 5N) are young stellar objects
probably embedded in the dense core SMM 4
(Nutter et al. 2005; Groppi et al. 2007).
IRS 5 as a whole is a Class I source
(Fig. 3; Wilking et al. 1997; Nutter et al. 2005; Hamaguchi et al. 2008).
However, Nisini et al. (2005) suggested
that IRS 5a may be a more evolved object, such as a T Tauri star,
on the basis of the lack of significant accretion activity.
It is not clear if IRS 5 has any outflow activity.
Groppi et al. (2007) reported a CO outflow in this region
though the driving source could not be specified clearly.
Subarcsecond-resolution imaging in the millimeter or submillimeter band
is necessary to understand the situation more clearly.

\subsubsection{IRS 5b}

The flux variability of IRS 5b needs a detailed discussion.
Though more extensive monitoring is necessary
to understand the flare activities,
it may be useful to consider two phases:
quiescent phase and active phase, depending on the total flux.
We suggest a flux threshold of 0.8 mJy for IRS 5b,
which is somewhat arbitrary.
The flux density in the quiescent phase (Tr 1, 3, 4, 5, 6, and 13)
is about 0.5 mJy on average.

Examining the light curve shown in Figure 13,
two periods of active phase can be recognized:
a short one in 1997 January (Tr 2)
and a long one in 1998 September--October (Tr 7--12).
Detailed information on the 1997 flare activity is not available,
and the duration of the active phase must have been less than 22 days.
The 1998 flare activity can be discussed in more detail.
The duration of the active phase was about 30 days,
and the flux fluctuated wildly during this period.
The average flux density during this active phase
was higher than that of the quiescent phase by a factor of about 4.
The light curve shows three flux peaks (Tr 7, 9, and 11)
with peak-to-peak intervals of about 10 days,
but the flux might have fluctuated more rapidly
if the sampling interval had been shorter.

The time scale of the flare activity of IRS 5b is much longer than 
that of the radio bursts of the Sun or flare stars,
typically a few minutes or shorter.
This month-scale flare activity of IRS 5b implies
that the cause of the flare
is not a local magnetic activity on the surface of the (proto)star
but probably a reconnection event
of global magnetic fields around the star
or even magnetic fields connecting the star and the circumstellar disk.
More details of the flare activity will be discussed in Paper II.

\subsubsection{IRS 5a}

IRS 5a is usually weak ($\sim$0.13 mJy) or undetectable,
but showed at least two active periods during the monitoring.
(Here, we set the quiescent-active phase threshold at 0.2 mJy.)
In 1998 June (Tr 4), IRS 5a went through an outburst
with the flux higher than the quiescent level by a factor of $\sim$15.
It is not clear how long was the duration of this outburst.

\input{fig14.tex}

In 1998 September--October (Tr 7--11),
the flux of IRS 5a increased mildly
and stayed at a level about twice the quiescent level.
This increase of flux is not an artifact of the image fitting process
used in measuring the flux of each binary component.
If the flux of IRS 5a had been at the quiescent level,
it would have been barely noticeable in the images,
but IRS 5a is clearly recognizable in the images during this active period.
For example, compare the Tr 9 image with the Tr 12 image:
the source structures in these two images are clearly different
even though the beam sizes are almost identical,
which clearly shows that the flux of IRS 5a
was significantly higher during Tr 9 than during Tr 12.

\subsubsection{Concurrent Activity in 1998}

What is really interesting
about the flare activity in 1998 September--October
is that both objects in the binary system are active at the same time.
If this concurrency is not a chance coincidence
but has a real physical reason,
there are three possibilities.
First, the binary companions may be interacting directly
through a physical process unknown so far.
Second, what was observed may be a giant magnetic structure
such as a circumbinary halo of ionized gas.
Third, they may be responding to an external cause,
such as an enhanced accretion of material from the envelope.

The possibility of the interacting (proto)binary
requires an underlying physical mechanism that is unknown so far.
If the intense flare activity of IRS 5b
was the cause of the flux enhancement of IRS 5a,
the light curves indicate
that such an interaction should happen in a time scale of several days.
The projected separation between IRS 5a and 5b is $\sim$0.9 light-day.
Then the order-of-magnitude velocity of the interaction agent
must be about 10\% of the speed of light or faster.
Such a high velocity is not completely unreasonable
because the circularly polarized radio emission of IRS 5b
implies the existence of mildly relativistic electrons.
Intrabinary magnetic fields between stars are often suggested
to explain the activity of stellar systems
such as RS CVn binaries (Dulk 1985),
and interacting helmet streamers were observed
around a young binary system (Massi et al. 2008).
The size of the IRS 5 system is probably too large
to have directly interacting stellar magnetic fields.
However, large-scale magnetic fields
involving circumstellar disk(s) and/or outflows
might provide a new mechanism of interaction.

Circumbinary halos and large-scale magnetic structures
are often suggested to explain the radio activity
of RS CVn binaries and related objects,
and some halos were imaged using the VLBI technique
(Mutel et al. 1985; Phillips et al. 1991).
It is unknown if a circumbinary halo can be as large as the IRS 5 system.
In principle, this possibility can be tested by imaging IRS 5
with an angular resolution of $\sim$0.2$''$ during an active phase.

In the third possibility,
the responses of IRS 5a and 5b can be synchronized
down to the crossing time of the accreting material.
The velocity of accreting gas can be approximated
by the free-fall velocity, $v_{\rm ff} \approx (2GM/r)^{1/2}$.
Assuming $M = 1 M_\odot$ and taking $r$ to be the binary separation,
we get $v_{\rm ff} \approx$ 3 km s$^{-1}$.
Then the crossing time is $\sim$200 yr,
which is too long to explain the concurrent activities of IRS 5a/b.
Therefore, the third possibility is improbable
unless there is a very fast triggering mechanism.

\subsection{B 5}

\input{fig15.tex}

\input{fig16.tex}

B 5 is a variable radio source (Fig. 2; Suters et al. 1996).
Figure 14 shows the radio continuum maps of B 5.
While the 3.5 cm map of Tr 1 did not have enough sensitivity
to see the source structure,
the 6.2 cm map of Tr 1 shows that B 5 is extended
along the northeast-southwest direction (P.A. = 44\arcdeg)
with a deconvolved size (FWHM) of 1.3$''$.
In the 3.5 cm map of Tr 13,
B 5 appears slightly extended along a similar direction (P.A. = 53\arcdeg)
with a deconvolved size of $\lesssim$0.6$''$.
The proper motion measurement of B 5 is unreliable
because the source is extended.
The peak position of the 6.2 cm source
has a small offset (0.6$''$ toward northeast)
with respect to the position of the 3.5 cm source.

The spectral index was highly negative (--1.3) during Tr 1
but positive (1.1) during Tr 13/14 (Table 3).
This difference is due to the high 3.5 cm flux during Tr 13,
which probably means
that B 5 went through an outburst during Tr 13.
This outburst was accompanied with circularly polarized emission,
suggesting magnetospheric origin (Paper II).
The centimeter flux during Tr 1
was obviously from a nonthermal emission process.
Such a large fluctuation of spectral index
was also reported by Suters et al. (1996)
based on monitoring observations in 1992.
However, the spectral index was never highly negative
in the 1992 observations,
and Suters et al. concluded
that the fluctuation was caused by an optical depth effect
of thermal free-free emission.
Miettinen et al. (2008) reported a negative spectral index,
confirming the nonthermal nature of the emission from B 5.
The complicated variability of the centimeter emission from B 5
may be caused by a combined effect of nonthermal emission process
and frequent flare events.

The nature of B 5 is an enigma.
Suters et al. (1996) argued against an extragalactic source
on the basis of the rapid variability.
The source size also supports this argument
because the radio flux from such an extended source cannot vary rapidly
if it is extragalactic.
Examinations of archival {\it Spitzer} images
did not produce any positive detection at the position of B 5.
The nondetection in the mid-IR or millimeter continuum
suggests that it is not a deeply embedded object
if located at the distance of the CrA cloud.
Feigelson et al. (1998) suggested B 5 to be a brown dwarf
based on a faint source in the near-IR images of Wilking et al. (1997),
but this region is confusing in optical and near-IR images
because of the nebulosity around IRS 2.

\subsection{IRS 1}

IRS 1 (HH 100 IR) is a point-like source,
and its flux is relatively stable.
Figure 15 shows the centimeter continuum maps toward the IRS 1 region.
The spectral index is positive
and consistent with optically thick thermal emission,
which agrees with the conclusion of Suters et al. (1996).
IRS 1 probably showed a proper motion
of 21 $\pm$ 10 mas yr$^{-1}$ toward the east,
but more observations are needed to confirm this value.

The 3.5 cm peak position agrees well
with the position of the mid-IR source IRAC 7 (Groppi et al. 2007).
IRS 1 is a variable X-ray source (Forbrich et al. 2006),
which suggests that the low variability of the radio flux
is probably due to the large optical depth in the centimeter band.
IRS 1 is a Class I object embedded in SMM 3,
and most of its bolometric luminosity is due to accretion
(Fig. 3; Wilking et al. 1992; Nisini et al. 2005;
Nutter et al. 2005; Groppi et al. 2007).
It is the driving source of HH 100
and probably other Herbig-Haro objects
(Strom et al. 1974; Hartigan \& Graham 1987; Wang et al. 2004).

\subsection{IRS 2}

IRS 2 is a point-like source,
and Figure 16 shows the radio continuum maps.
IRS 2 is located outside the field of view in our 3.5 cm observations,
but is within the field of view of the 3.5 cm observations
of projects AC 481 and AM 596.
The 3.5 cm flux density varies between 0.25 and 0.64 mJy.
The spectral index between 3.5 and 6.2 cm,
estimated from the average flux densities, is 1.3 $\pm$ 0.7,
but this value is highly uncertain
because there was no observation
made simultaneously in both wavelength bands.
The positive spectral index implies
that thermal free-free radiation
may be the emission mechanism at the centimeter band.

\input{fig17.tex}

\input{fig18.tex}

The 3.5 cm peak position agrees well
with the near-IR source position (Fig. 16).
IRS 2 is a strong and variable X-ray source (Forbrich et al. 2006).
IRS 2 seems to be embedded in SMM 5
and shows evidences of active accretion
(Nisini et al. 2005; Nutter et al. 2005).
The bolometric temperature and luminosity of IRS 2,
calculated using the infrared fluxes
from {\it Spitzer} archival data (Table 4),
are 457 $\pm$ 17 K and 10.0 $\pm$ 0.1 $L_\odot$, respectively.
This temperature confirms that IRS 2 is a Class I protostar
as previously suggested by Wilking et al. (1992).
The IRAC color-color diagram also places IRS 2
inside the Class 0/I domain (Fig. 3).

\subsection{R CrA}

R CrA is a point-like source,
and Figure 17 shows the radio continuum maps.
While R CrA is dominating this region in the optical and near-IR bands,
it is relatively weak in radio.
R CrA is detectable in all the data sets analyzed
and shows some flux variability (Fig. 2).
The spectral index of the centimeter continuum is positive,
suggesting that the radio flux comes from optically thick thermal emission.
The proper motion of R CrA is consistent
with that of the R CrA association (see \S~4.4).

The 3.5 cm peak position agrees well
with the near-IR source position (Fig. 17).
It is not clear how strong the submillimeter emission from R CrA is
(Groppi et al. 2007).
R CrA is a Herbig Ae star of $\sim$3.5 $M_\odot$,
and it is also an interesting X-ray source
with a plasma temperature of $\sim$100 MK (Forbrich et al. 2006).
To explain the X-ray emission,
Forbrich et al. (2006) suggested that R CrA may be a binary system:
an optically bright Herbig Ae star and an X-ray bright Class I source.
Takami et al. (2003) also suggested that R CrA may be a binary system
based on spectro-astrometric observations,
but, in this case, both members must be visible in the H$\alpha$ line.

\subsection{T CrA}

Radio emission from T CrA has not been reported before,
but it was detected in our maps as a weak point-like source.
Careful examinations of archival data revealed
that T CrA was detectable most of the time.
Figure 18 shows the radio continuum maps,
and Table 6 lists the flux densities.
The spectral index is negative,
but the uncertainty is too large to constrain the emission mechanism.

\input{tab6.tex}

The 3.5 cm peak position agrees well
with the position of the mid-IR source IRAC 1 (Groppi et al. 2007).
T CrA is a weak X-ray source (Forbrich et al. 2007)
and was not detected in the submillimeter continuum (Groppi et al. 2007).
T CrA is a Herbig star of type F0e.
Takami et al. (2003) suggested that T CrA may be a binary system.
Wang et al. (2004) suggested
that T CrA may be the driving source of the Herbig-Haro object HH 733.

\subsection{WMB 55}

\input{fig19.tex}

Centimeter continuum emission from WMB 55 has not been reported before,
but it was detected in our 3.5 cm map as a weak point-like source.
Examinations of archival data revealed
that the 3.5 cm flux of WMB 55 was detectable in a few observing runs,
but it was not detected in the 6.2 cm continuum.
Table 6 lists the flux densities,
and Figure 19 shows the 3.5 cm continuum maps.
The nondetection in the 6.2 cm continuum implies
that the spectral index is positive,
but the uncertainty is too large to constrain the emission mechanism.

The 3.5 cm peak position agrees well
with the position of the mid-IR source IRAC 3 (Groppi et al. 2007).
WMB 55 was not detected in the X-ray band.
WMB 55 is a Class I source embedded in SMM 2
(Fig. 3; Chini et al. 2003; Nutter et al. 2005).
Wang et al. (2004) suggested
that WMB 55 is a probable driving source
of the Herbig-Haro objects HH 733 and/or HH 734.
HH 104 A/B are also located in the vicinity of WMB 55,
but it is not clear if they are related with WMB 55.

\subsection{IRS 6}

IRS 6 was not detected in our maps,
though Forbrich et al. (2006) reported
that it was detected when several data sets were combined.
Surprisingly, examinations of archival data revealed
that IRS 6 went through an outburst on 1998 October 13 (Tr 12)
while it was undetectable or only marginally detectable in the other epochs.
Miettinen et al. (2008) reported another outburst detected in 1998 January.

\input{fig20.tex}

Figure 20 shows the 3.5 cm continuum map at the epoch of the outburst.
The source was slightly elongated
along the northwest-southeast direction (P.A. $\approx$ --41\arcdeg)
with a deconvolved size of 0.7$''$,
and the total flux density was 0.44 $\pm$ 0.05 mJy.
Since IRS 6 was undetected during Tr 11,
the onset time-scale was less than 3 days.
It is not clear how long was the duration of this outburst
because Tr 12 was the last observing track of the 1998 observing campaign.

The peak position at the outburst
agrees well, within 0.2$''$, with the X-ray source position,
which suggests that the same source is responsible
for both the radio and the X-ray emission.
(The X-ray source position was determined
using the images discussed in Hamaguchi et al. 2008.)
Since IRS 6 is a near-IR binary system (Nisini et al. 2005),
the radio/X-ray source may be one of the binary members.
Comparing the near-IR source positions with the radio source position,
after applying the astrometric correction
determined using IRS 5a/b (see \S~5.3),
the radio source can be identified as IRS 6a.
Interestingly, the mid-IR source IRAC 8
is located halfway between IRS 6a and 6b,
suggesting that the mid-IR fluxes of the binary members are comparable.
X-ray emission was never detected at the position of IRS 6b.
In contrast to the IRS 5 system,
only one member of the IRS 6 system
seems to show detectable magnetic activities.

IRS 6a showed little sign of accretion activity (Nisini et al. 2005)
and was suggested to be a T Tauri star (Forbrich et al. 2006).
However, the spectral class of IRS 6 was unclear
because a detailed SED was not available.
The 850 $\mu$m continuum map of Groppi et al. (2007)
does show an unnumbered local peak at the position of IRS 6,
which indicates that IRS 6 may be an embedded object.
The bolometric temperature and luminosity of IRS 6,
calculated using the infrared fluxes
from the {\it Spitzer} archival data (Table 4),
are 424 $\pm$ 12 K and $\sim$0.4 $L_\odot$, respectively.
This temperature suggests that IRS 6 is a Class I protostar.
The IRAC color-color diagram also places IRS 6
in the Class 0/I domain (Fig. 3).
Wang et al. (2004) suggested
that IRS 6 may be driving a giant Herbig-Haro flow:
HH 730, HH 860, HH 104 C/D, and HH 99.

\section{SUMMARY}

We observed the R CrA region using the VLA in the 3.5 and 6.2 cm continuum
to image the young stellar objects
and study the star forming activities in this region.
Archival VLA data from recent observations
made in extended configurations were also analyzed.
A total of 16 sources was detected.
The source properties were studied by investigating source structure,
flux variability, spectral slope, and proper motion
and also by comparing with the sources detected in shorter wavelength.
The main results are summarized as follows:

1.
The centimeter emission of IRS 7A traces the thermal radio jet
that seems to be flowing in the north-south direction,
and the outflow activity has become enhanced in recent years.
A recently ejected outflow knot was found to be moving
at a speed of 100 km s$^{-1}$ or higher.
The main peak position of the radio emission
agrees with the mid-IR and X-ray positions,
which suggests that the X-ray emission comes
from the central (proto)star or a very vicinity of it.

2.
The Class 0 source SMA 2 is associated
with the centimeter continuum sources B 9a/b
and the 7 mm continuum source CT 2.
While the detailed structure of the B 9 complex needs further studies,
it may be a multiple-protostar system
that is driving at least two outflows (FPM 10 and FPM 13).
The negative spectral slope of B 9a/b indicates
that the radiation process is nonthermal.

3.
The centimeter continuum images of IRS 7B show
a combination of a compact source and an extended structure.
The compact source corresponds to the Class 0/I source SMA 1.
The X-ray source position also agrees well with the radio peak position,
and IRS 7B (SMA 1) is one of the youngest known objects
that were detected in X-rays,
which suggests that energetic magnetic activities
start very early in the protostellar stage.
The extended structure is probably a bipolar outflow driven by IRS 7B.

4.
The Class I binary system IRS 5 was resolved in the 3.5 cm continuum maps.
The binary separation is 0.9$''$ (160 AU).
Both members of the system show radio flare activities and X-ray emission.
IRS 5 is one of the youngest known binary systems
that is showing magnetic activities in both members.

5.
The radio light curve of IRS 5b shows at least two flare activities,
and the duration of active phase is about a month.
Such a long time-scale suggests
that the flare activity involves large-scale magnetic fields.

6.
During the 1998 September--October flare event of IRS 5b,
the radio flux of IRS 5a also showed a significant enhancement.
This concurrent activity suggests a few possibilities:
IRS 5a and 5b are interacting directly,
they are surround by a circumbinary halo,
or they are responding to an external trigger.
IRS 5 may be the youngest known example of interacting binaries.

7.
The variable radio source B 5
was found to be an extended ($\sim$1$''$) object.
The spectral slope also varies significantly,
suggesting that both thermal and nonthermal radiation processes are working.
The rapid variability and the source size rule out an extragalactic source.
The nature of B 5, however, still remains unclear.

8.
The Class I sources IRS 1 and IRS 2 show
positive spectral slopes.
The centimeter emission of these sources
is likely coming from optically thick thermal free-free radiation.
The flux density of IRS 1 is relatively stable.

9. 
The Herbig Ae star R CrA is a relatively weak radio source,
and the spectral slope is consistent with thermal emission.
The marginally detected proper motion, 28 mas yr$^{-1}$ toward southeast,
is consistent with that of the R CrA association.

10.
Radio detections of the Herbig star T CrA and the Class I source WMB 55
are reported here for the first time.
The emission mechanism of these sources is not clear.

11.
The near-IR binary IRS 6 is
usually weak or undetectable in the centimeter continuum,
but a radio flare event was detected in 1998 October.
The radio source positionally coincides with the X-ray source
and was tentatively identified as IRS 6a.
Only one member of the IRS 6 binary system
seems to show detectable magnetic activities.
Infrared data suggests that IRS 6 is a Class I source.

\acknowledgements

We thank K.-H. Kim for helpful discussions.
This work was supported by the LRG Program of KASI.
K. H. is supported by the NASA Astrobiology Program under CAN 03-OSS-02.
J.-E. L. gratefully acknowledges the support
by the Korea Science and Engineering Foundation (KOSEF)
under a cooperative agreement with the Astrophysical Research Center
for the Structure and Evolution of the Cosmos (ARCSEC).
The National Radio Astronomy Observatory is
a facility of the National Science Foundation
operated under cooperative agreement by Associated Universities, Inc.
This work is based in part on observations
made with the {\it Spitzer} Space Telescope,
which is operated by the Jet Propulsion Laboratory,
California Institute of Technology under a contract with NASA.
This publication makes use of data products
from the Two Micron All Sky Survey,
which is a joint project of the University of Massachusetts
and the Infrared Processing
and Analysis Center/California Institute of Technology,
funded by the National Aeronautics and Space Administration
and the National Science Foundation.

\enlargethispage{-14\baselineskip}


\end{document}

%% file: tab1.tex
\begin{deluxetable}{p{16mm}cccrccrc}
\tabletypesize{\small}
\tablecaption{Parameters of the VLA Observing Runs}%
\tablewidth{0pt}
\tablehead{
&&& \multicolumn{5}{c}{\sc Synthesized Beam} \\
\cline{4-8}
&& \colhead{\sc Frequency} & \multicolumn{2}{c}{Natural weight}
&& \multicolumn{2}{c}{Uniform weight} & \colhead{\sc Array}\\
\cline{4-5} \cline{7-8}
\colhead{\sc Track} & \colhead{\sc Date} & \colhead{(GHz)}
& \colhead{Size} & \colhead{P.A.} && \colhead{Size} & \colhead{P.A.}
& \colhead{\sc Configuration} }%
\startdata
Tr 1\dotfill  & 1996. 12. 29. & 4.9 & 1.21 $\times$ 0.50 &  $-$6.5 && 0.86 $\times$ 0.35 &  $-$7.1 & A \\
              &               & 8.5 & 0.70 $\times$ 0.31 &  $-$2.5 && 0.51 $\times$ 0.20 &     0.1 & A \\
Tr 2\dotfill  & 1997. 01. 19. & 8.5 & 0.82 $\times$ 0.59 &    10.6 && 0.57 $\times$ 0.20 &     0.8 & BnA \\
Tr 3\dotfill  & 1997. 01. 20. & 8.5 & 1.04 $\times$ 0.89 & $-$10.4 && 0.84 $\times$ 0.22 &    10.6 & BnA \\
Tr 4\dotfill  & 1998. 06. 27. & 8.5 & 1.13 $\times$ 0.69 &    23.8 && 0.82 $\times$ 0.52 &    41.0 & BnA \\
Tr 5\dotfill  & 1998. 07. 19. & 8.5 & 2.61 $\times$ 0.78 &  $-$1.9 && 2.38 $\times$ 0.54 &  $-$0.8 & B \\
Tr 6\dotfill  & 1998. 09. 07. & 8.5 & 2.61 $\times$ 0.78 &     5.4 && 2.37 $\times$ 0.54 &     4.7 & B \\
Tr 7\dotfill  & 1998. 09. 19. & 8.5 & 2.54 $\times$ 0.80 &     0.0 && 2.30 $\times$ 0.53 &     0.4 & B \\
Tr 8\dotfill  & 1998. 09. 27. & 8.5 & 2.71 $\times$ 0.79 &    11.6 && 2.43 $\times$ 0.52 &    11.8 & B \\
Tr 9\dotfill  & 1998. 10. 02. & 8.5 & 2.58 $\times$ 0.79 &  $-$5.4 && 2.25 $\times$ 0.54 &  $-$4.6 & B \\
Tr 10\dotfill & 1998. 10. 06. & 8.5 & 2.65 $\times$ 0.79 &  $-$6.6 && 2.43 $\times$ 0.54 &  $-$5.2 & B \\
Tr 11\dotfill & 1998. 10. 10. & 8.5 & 2.54 $\times$ 0.79 &     0.1 && 2.30 $\times$ 0.53 &     0.4 & B \\
Tr 12\dotfill & 1998. 10. 13. & 8.5 & 2.60 $\times$ 0.80 &  $-$5.6 && 2.25 $\times$ 0.54 &  $-$4.6 & B \\
Tr 13\dotfill & 2005. 02. 03. & 8.5 & 1.01 $\times$ 0.74 &    23.4 && 0.73 $\times$ 0.54 &    33.9 & BnA \\
Tr 14\dotfill & 2005. 02. 12. & 4.9 & 2.10 $\times$ 1.35 &     9.9 && 1.45 $\times$ 1.02 &    14.7 & BnA \\
\enddata\\
\tablecomments{Units of beam size and P.A.
               are arcseconds and degrees, respectively.
               Tracks Tr 2--3 correspond to the observing runs
               presented by Feigelson et al. 1998.
               Tracks Tr 4--12 correspond to epochs R1--9
               in Table 1 of Forbrich et al. 2006.}%
\end{deluxetable}

%% file: tab2.tex
\begin{deluxetable}{lccl}
\tabletypesize{\small}
\tablecaption{Centimeter Continuum Sources in the R CrA Region}%
\tablewidth{0pt}
\tablehead{
& \multicolumn{2}{c}{\sc Peak Position\tablenotemark{a}} \\
\cline{2-3}
\colhead{\sc Source}
& \colhead{$\alpha_{\rm J2000.0}$} & \colhead{$\delta_{\rm J2000.0}$}
& \colhead{\sc Associated Objects\tablenotemark{b}} }%
\startdata
1\dotfill  & 19 01 41.58 & --36 58 31.0 & IRS 2, FCW 1, SMM 5 \\
2\dotfill  & 19 01 43.27 & --36 59 12.3 & B 5 \\
3\dotfill  & 19 01 48.02 & --36 57 22.5 & IRS 5a, B 7, SMM 4, IRAC 10 \\
4\dotfill  & 19 01 48.06 & --36 57 21.8 & IRS 5b, B 7, SMM 4 \\
5\dotfill  & 19 01 50.46 & --36 56 37.6 & IRS 6, FPM 6, IRAC 8 \\
6\dotfill  & 19 01 50.67 & --36 58 09.7 & IRS 1, B 8, SMM 3, IRAC 7 \\
7\dotfill  & 19 01 53.68 & --36 57 08.2 & R CrA, FCW 5, IRAC 6 \\
8\dotfill  & 19 01 55.04 & --36 57 16.6 & FPM 10 \\
9\dotfill  & 19 01 55.26 & --36 57 16.9 & B 9, CT 2, SMM 1C, SMA 2 \\
10\dotfill & 19 01 55.30 & --36 57 16.7 & B 9, CT 2, SMM 1C, SMA 2 \\
11\dotfill & 19 01 55.32 & --36 57 22.1 & IRS 7A, CT 4, IRAC 5 \\
12\dotfill & 19 01 55.34 & --36 57 12.9 & FPM 13 \\
13\dotfill & 19 01 55.61 & --36 57 43.4 & SMM 1A \\
14\dotfill & 19 01 56.41 & --36 57 28.1 & IRS 7B, SMM 1B, SMA 1, IRAC 4 \\
15\dotfill & 19 01 58.55 & --36 57 08.5 & WMB 55, SMM 2, IRAC 3 \\
16\dotfill & 19 01 58.78 & --36 57 49.9 & T CrA, IRAC 1 \\
\enddata\\
\tablecomments{Units of right ascension are hours, minutes, and seconds,
               and units of declination are degrees, arcminutes,
               and arcseconds.}%
\tablenotetext{a}{For most sources, the positions are
                  from the natural-weight 3.5 cm continuum map of Tr 1.
                  For source 1 (IRS 2),
                  the position is from the 6.2 cm map of Tr 1.
                  For source 5 (IRS 6),
                  the position is from the 3.5 cm map of Tr 12.
                  For sources 8, 12, and 15 (FPM 10, FPM 13, and WMB 55),
                  the positions are from the 3.5 cm map of Tr 13.
                  For source 13,
                  the position is from the 6.2 cm map of Tr 14.}%
\tablenotetext{b}{Source numbers from IRS: Taylor \& Storey 1984,
                  B: Brown 1987, WMB: Wilking et al. 1997,
                  FCW: Feigelson et al. 1998, CT: Choi \& Tatematsu 2004,
                  SMM: Nutter et al. 2005 and Groppi et al. 2007,
                  FPM: Forbrich et al. 2006,
                  and SMA/IRAC: Groppi et al. 2007.}%
\end{deluxetable}

%% file: fig1.tex
\begin{figure*}[p]
\centerline{\plotpanel{FChart.X.eps}{145mm}{1}}
\vspace*{5mm}
\centerline{\plotpanel{FChart.XC.eps}{145mm}{1}}
\caption{\small\baselineskip=0.825\baselineskip
Finding charts of the radio continuum sources in the R CrA region.
($a$)
Map of the $\lambda$ = 3.5 cm continuum.
This 3.5 cm map was made by averaging all the images from tracks Tr 1--13.
For each track, a natural-weight image was made.
Each of these images was corrected for the primary beam response
and convolved to an angular resolution of 3$''$,
and then all the 13 images were averaged.
The contour levels are 1, 2, 4, 8, 16, 32, and 64 times 0.09 m\Jypb.
($b$)
Natural-weight maps of the $\lambda$ = 3.5 and 6.2 cm continuum
toward the IRS 7 region observed in 2005 February (Tr 13/14).
The contour levels are 1, 2, 4, 8, 16, 32, and 64 times 0.06 m\Jypb.
Dashed contours are for negative levels.
The rms noise is 0.02 m\Jypb.
Shown in the bottom right corner are the synthesized beams.
The straight line near the bottom left corner
corresponds to 1000 AU at a distance of 170 pc.
{\it Plus signs}:
Submillimeter continuum sources SMM 1C, SMM 1B, and SMM 1A,
from north to south (Groppi et al. 2007).}
\end{figure*}

%% file: tab3.tex
\begin{deluxetable}{p{15mm}lcccccl@{~$\pm$~}l}
\tabletypesize{\small}
\tablecaption{Flux Densities of the Radio Sources in the R CrA Region}%
\tablewidth{0pt}
\tablehead{
&& \multicolumn{2}{c}{$F$(6.2 cm)} && \multicolumn{2}{c}{$F$(3.5 cm)} \\
\cline{3-4} \cline{6-7}
\colhead{\sc Source} & \colhead{\sc Name}
& \colhead{Peak} & \colhead{Total} && \colhead{Peak} & \colhead{Total}
& \multicolumn{2}{c}{$\alpha$\tablenotemark{a}} }%
\startdata
\multicolumn{9}{c}{Tr 1: 1996 December} \\
\cline{1-9}
1\dotfill  & IRS 2\tablenotemark{b}
                     & 0.15 $\pm$ 0.03 & 0.25 $\pm$ 0.04 && \nodata         & \nodata         & \multicolumn{2}{c}{\nodata} \\
2\dotfill  & B 5     & 0.46 $\pm$ 0.03 & 1.06 $\pm$ 0.11 && 0.42 $\pm$ 0.04 & 0.51 $\pm$ 0.07 &  $-$1.3  & 0.3 \\
3\dotfill  & IRS 5a\tablenotemark{c}
                     & \nodata         & \nodata         && 0.11 $\pm$ 0.02 & 0.12 $\pm$ 0.02 & \multicolumn{2}{c}{\nodata} \\
4\dotfill  & IRS 5b\tablenotemark{c}
                     & \nodata         & \nodata         && 0.17 $\pm$ 0.02 & 0.25 $\pm$ 0.02 & \multicolumn{2}{c}{\nodata} \\
3+4\dotfill & IRS 5\tablenotemark{c}
                     & 0.22 $\pm$ 0.02 & 0.43 $\pm$ 0.03 && \nodata         & 0.35 $\pm$ 0.07 &  $-$0.4  & 0.4 \\
5\dotfill  & IRS 6   & \nodata         & \nodata         && \nodata         & \nodata         & \multicolumn{2}{c}{\nodata} \\
6\dotfill  & IRS 1   & 0.16 $\pm$ 0.02 & 0.16 $\pm$ 0.02 && 0.22 $\pm$ 0.02 & 0.26 $\pm$ 0.03 & \phs0.9  & 0.3 \\
7\dotfill  & R CrA   & 0.10 $\pm$ 0.02 & 0.11 $\pm$ 0.04 && 0.19 $\pm$ 0.02 & 0.19 $\pm$ 0.02 & \phs1.0  & 0.7 \\
8\dotfill  & FPM 10  & \nodata         & \nodata         && 0.08 $\pm$ 0.02 & 0.11 $\pm$ 0.02 & \multicolumn{2}{c}{\nodata} \\
9\dotfill  & B 9b\tablenotemark{d,e}
                     & 0.39 $\pm$ 0.03 & 0.32 $\pm$ 0.03 && 0.15 $\pm$ 0.02 & 0.19 $\pm$ 0.04 &  $-$0.9  & 0.4 \\
10\dotfill & B 9a\tablenotemark{d,e}
                     & 0.40 $\pm$ 0.03 & 1.09 $\pm$ 0.03 && 0.49 $\pm$ 0.02 & 0.90 $\pm$ 0.04 &  $-$0.35 & 0.09 \\
9+10\dotfill & B 9   & \nodata         & 1.33 $\pm$ 0.03 && \nodata         & 1.09 $\pm$ 0.04 &  $-$0.36 & 0.08 \\
11\dotfill & IRS 7A  & 2.61 $\pm$ 0.02 & 3.63 $\pm$ 0.07 && 2.40 $\pm$ 0.02 & 3.71 $\pm$ 0.07 & \phs0.04 & 0.05 \\
12\dotfill & FPM 13  & 0.10 $\pm$ 0.02 & 0.16 $\pm$ 0.03 && 0.08 $\pm$ 0.02 & 0.14 $\pm$ 0.02 &  $-$0.2  & 0.4 \\
13\dotfill & CHLT 13 & \nodata         & \nodata         && \nodata         & \nodata         & \multicolumn{2}{c}{\nodata} \\
14\dotfill & IRS 7B  & 0.92 $\pm$ 0.02 & 2.54 $\pm$ 0.11 && 0.73 $\pm$ 0.02 & 1.16 $\pm$ 0.03 &  $-$1.41 & 0.09 \\
15\dotfill & WMB 55  & \nodata         & \nodata         && \nodata         & \nodata         & \multicolumn{2}{c}{\nodata} \\
16\dotfill & T CrA   & 0.13 $\pm$ 0.02 & 0.13 $\pm$ 0.02 && 0.10 $\pm$ 0.02 & 0.10 $\pm$ 0.02 &  $-$0.5  & 0.5 \\
\cline{1-9}
\multicolumn{9}{c}{Tr 13/14: 2005 February} \\
\cline{1-9}
1\dotfill  & IRS 2\tablenotemark{b}
                     & 0.17 $\pm$ 0.03 & 0.17 $\pm$ 0.03 && \nodata         & \nodata         & \multicolumn{2}{c}{\nodata} \\
2\dotfill  & B 5     & 1.06 $\pm$ 0.03 & 1.21 $\pm$ 0.04 && 1.78 $\pm$ 0.05 & 2.18 $\pm$ 0.14 & \phs1.06 & 0.13 \\
3\dotfill  & IRS 5a\tablenotemark{c,e}
                     & \nodata         & \nodata         && 0.16 $\pm$ 0.03 & 0.16 $\pm$ 0.03 & \multicolumn{2}{c}{\nodata} \\
4\dotfill  & IRS 5b\tablenotemark{c,e}
                     & \nodata         & \nodata         && 0.41 $\pm$ 0.03 & 0.41 $\pm$ 0.03 & \multicolumn{2}{c}{\nodata} \\
3+4\dotfill & IRS 5\tablenotemark{c}
                     & 0.58 $\pm$ 0.02 & 0.77 $\pm$ 0.03 && \nodata         & 0.77 $\pm$ 0.05 & \phs0.00 & 0.14 \\
5\dotfill  & IRS 6   & \nodata         & \nodata         && \nodata         & \nodata         & \multicolumn{2}{c}{\nodata} \\
6\dotfill  & IRS 1   & 0.27 $\pm$ 0.02 & 0.30 $\pm$ 0.04 && 0.46 $\pm$ 0.02 & 0.50 $\pm$ 0.05 & \phs0.9  & 0.3 \\
7\dotfill  & R CrA   & 0.15 $\pm$ 0.02 & 0.15 $\pm$ 0.02 && 0.28 $\pm$ 0.02 & 0.28 $\pm$ 0.02 & \phs1.1  & 0.3 \\
8\dotfill  & FPM 10  & 0.27 $\pm$ 0.02 & 0.42 $\pm$ 0.03 && 0.19 $\pm$ 0.02 & 0.30 $\pm$ 0.04 &  $-$0.6  & 0.3 \\
9+10\dotfill & B 9\tablenotemark{f}
                     & 1.27 $\pm$ 0.02 & 1.76 $\pm$ 0.07 && 1.14 $\pm$ 0.02 & 1.56 $\pm$ 0.04 &  $-$0.22 & 0.09 \\
11\dotfill & IRS 7A  & 5.08 $\pm$ 0.02 & 6.24 $\pm$ 0.08 && 4.22 $\pm$ 0.02 & 6.09 $\pm$ 0.06 &  $-$0.04 & 0.03 \\
12\dotfill & FPM 13  & 0.31 $\pm$ 0.02 & 0.47 $\pm$ 0.04 && 0.16 $\pm$ 0.02 & 0.52 $\pm$ 0.06 & \phs0.2  & 0.3 \\
13\dotfill & CHLT 13 & 0.17 $\pm$ 0.02 & 0.17 $\pm$ 0.02 && \nodata         & \nodata         & \multicolumn{2}{c}{\nodata} \\
14\dotfill & IRS 7B  & 0.89 $\pm$ 0.02 & 2.56 $\pm$ 0.12 && 0.81 $\pm$ 0.02 & 2.31 $\pm$ 0.15 &  $-$0.19 & 0.14 \\
15\dotfill & WMB 55  & \nodata         & \nodata         && 0.15 $\pm$ 0.02 & 0.17 $\pm$ 0.03 & \multicolumn{2}{c}{\nodata} \\
16\dotfill & T CrA   & \nodata         & \nodata         && 0.13 $\pm$ 0.02 & 0.17 $\pm$ 0.02 & \multicolumn{2}{c}{\nodata} \\
\enddata\\
\tablecomments{Flux densities are corrected for the primary beam response,
               and the unit is mJy (beam$^{-1}$).
               For most sources, the flux densities are
               from the natural-weight maps.
               In most cases, total flux of each source
               was measured in a box,
               typically a circumscribed rectangle of 2$\sigma$ contour.}%
\tablenotetext{a}{Spectral index between 6.2 cm and 3.5 cm.
                  In 1996 December, the 6.2 cm and the 3.5 cm observations
                  were made in a single track.
                  In 2005 February, the 6.2 cm observations were made
                  9 days after the 3.5 cm observations,
                  and the spectral index of rapidly varying sources
                  should be interpreted with caution.}%
\tablenotetext{b}{IRS 2 is located outside the field of view
                  of the 3.5 cm maps.}%
\tablenotetext{c}{IRS 5a and 5b were not clearly separated
                  in the 6.2 cm maps.}%
\tablenotetext{d}{The 6.2 cm fluxes of B 9a and 9b
                  are from the uniform-weight maps.}%
\tablenotetext{e}{The 3.5 cm fluxes of these sources
                  are from the uniform-weight maps.}%
\tablenotetext{f}{B 9a and 9b were not clearly separated
                  in either the 6.2 cm or the 3.5 cm maps.}%
\end{deluxetable}

%% file: fig2.tex
\begin{figure*}
\ploteps{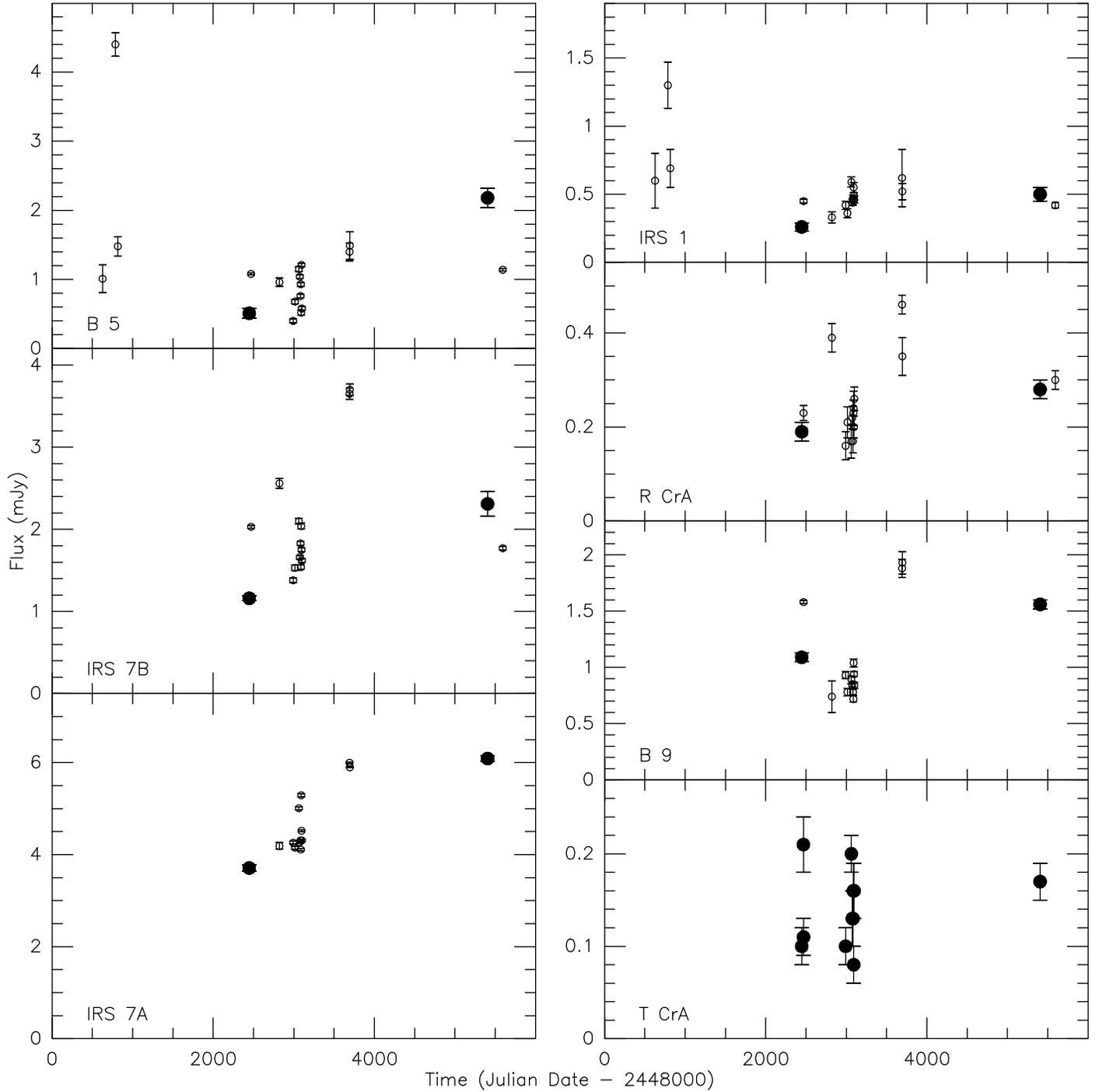}{185mm}
\caption{\small\baselineskip=0.825\baselineskip
Variations of 3.5 cm flux densities.
{\it Filled circles:}
Flux densities from this work.
{\it Open circles:}
Flux densities from previously published works
(Suters et al. 1996; Feigelson et al. 1998;
Forbrich et al. 2006, 2007; Miettinen et al. 2008).}
\end{figure*}

%% file: tab4.tex
\begin{deluxetable}{p{16mm}cccccc}
\tabletypesize{\small}
\tablecaption{Infrared Flux Densities}%
\tablewidth{0pt}
\tablehead{
\colhead{Source} & \colhead{$S$(3.6 $\mu$m)} & \colhead{$S$(4.5 $\mu$m)}
& \colhead{$S$(5.8 $\mu$m)} & \colhead{$S$(8 $\mu$m)}
& \colhead{$S$(24 $\mu$m)} & \colhead{$S$(70 $\mu$m)} }%
\startdata
IRS 2\dotfill & 2.57\phn & 3.65\phn & 6.95 & 9.61 &    17.4\phn &    35.7 \\
IRS 6\dotfill & 0.099    & 0.162    & 0.21 & 0.23 & \phn0.44    & \phn4.7 \\
\enddata\\
\tablecomments{Fluxes are from the {\it Spitzer} data,
               and the unit is Jy.
               Photometric uncertainties are $\sim$10\%.}%
\end{deluxetable}

%% file: fig3.tex
\begin{figure}
\ploteps{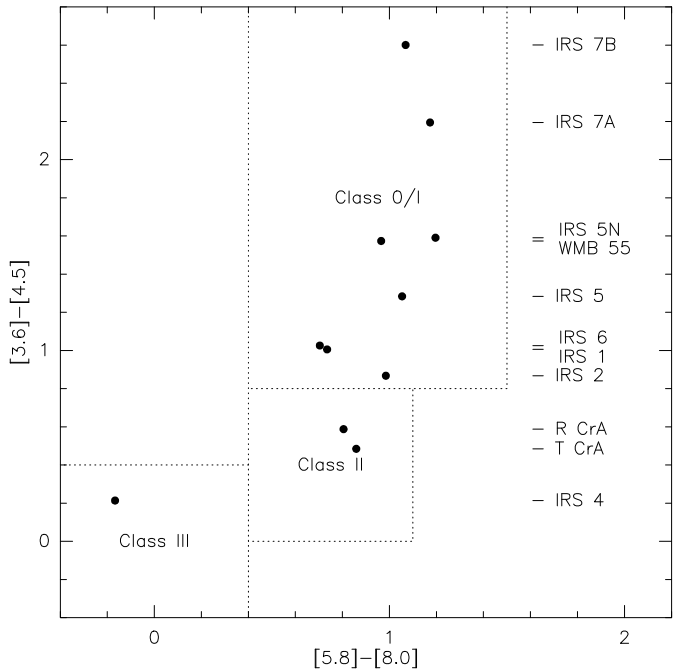}{88mm}
\caption{\small\baselineskip=0.825\baselineskip
Color-color diagram for the young stellar objects in the R CrA region
from the {\it Spitzer} IRAC 3.6, 4.5, 5.8, and 8.0 $\mu$m data.
For IRS 2 and IRS 6, the photometric data are listed in Table 4.
For the other sources, the flux densities are from Groppi et al. (2007).
The dotted boxes delineate the domains of Class 0/I, II, and III sources
(Allen et al. 2004; Lee et al. 2006).}
\end{figure}

%% file: fig4.tex
\begin{figure}
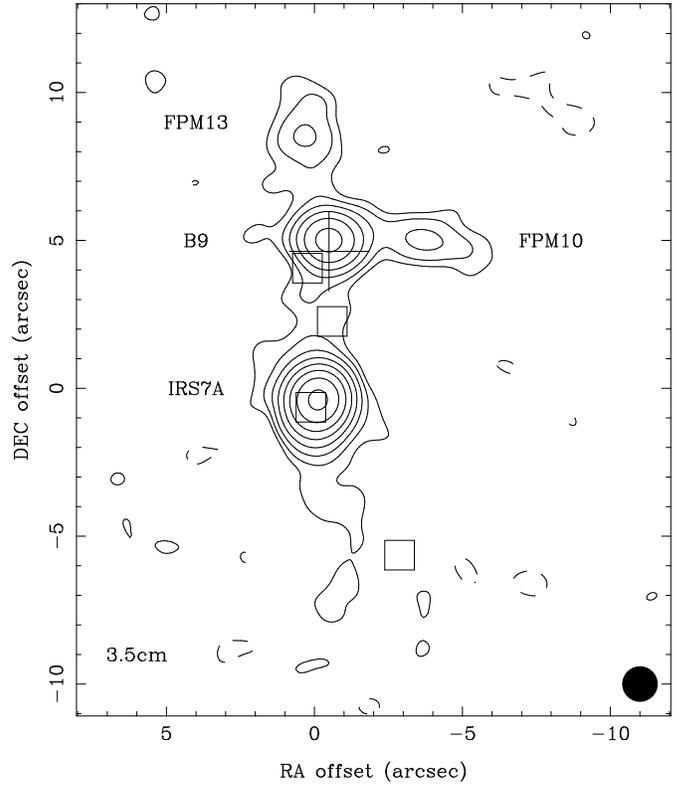

\ploteps{IRS7A.Xavg.eps}{88mm}
\caption{\small\baselineskip=0.825\baselineskip
Map of the 3.5 cm continuum toward the IRS 7A region
from high-resolution data sets: Tr 1, 2, 3, 4, and 13.
For each data set, a natural-weight image was made.
Each of these images was corrected for the primary beam response
and convolved to an angular resolution of 1.2$''$
(shown in the bottom right corner),
and then all the five images were averaged.
The contour levels are 1, 2, 4, 8, 16, 32, 64, and 128 times 0.025 m\Jypb.
Dashed contours are for negative levels.
{\it Squares}:
The 7 mm continuum sources CT 2, 3, 4, and 5, from north to south
(Choi \& Tatematsu 2004).
{\it Plus sign}:
The 1.1 mm continuum source SMA 2 (Groppi et al. 2007).}
\end{figure}

%% file: fig5.tex
\begin{figure}
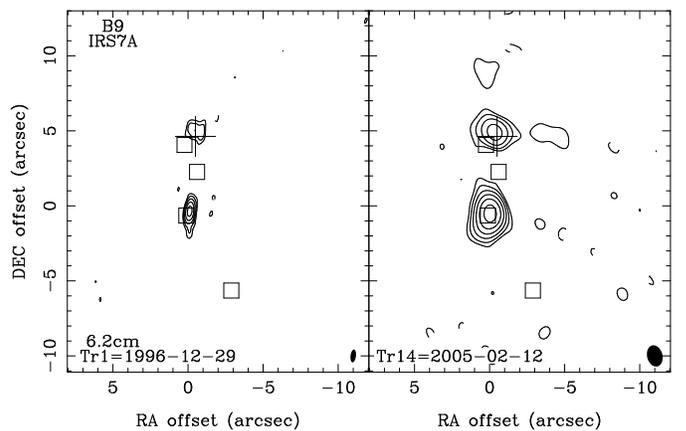

\ploteps{IRS7A.Cuni.eps}{88mm}
\caption{\small\baselineskip=0.825\baselineskip
Uniform-weight maps of the 6.2 cm continuum toward the IRS 7A region.
The contour levels are 1, 2, 4, 8, 16, and 32 times 0.10 m\Jypb.
Markers are the same as those in Fig. 4.
Observation dates and track numbers are labeled.
Shown in the bottom right corner are the synthesized beams (see Table 1).}
\end{figure}

%% file: fig6.tex
\begin{figure*}[p]
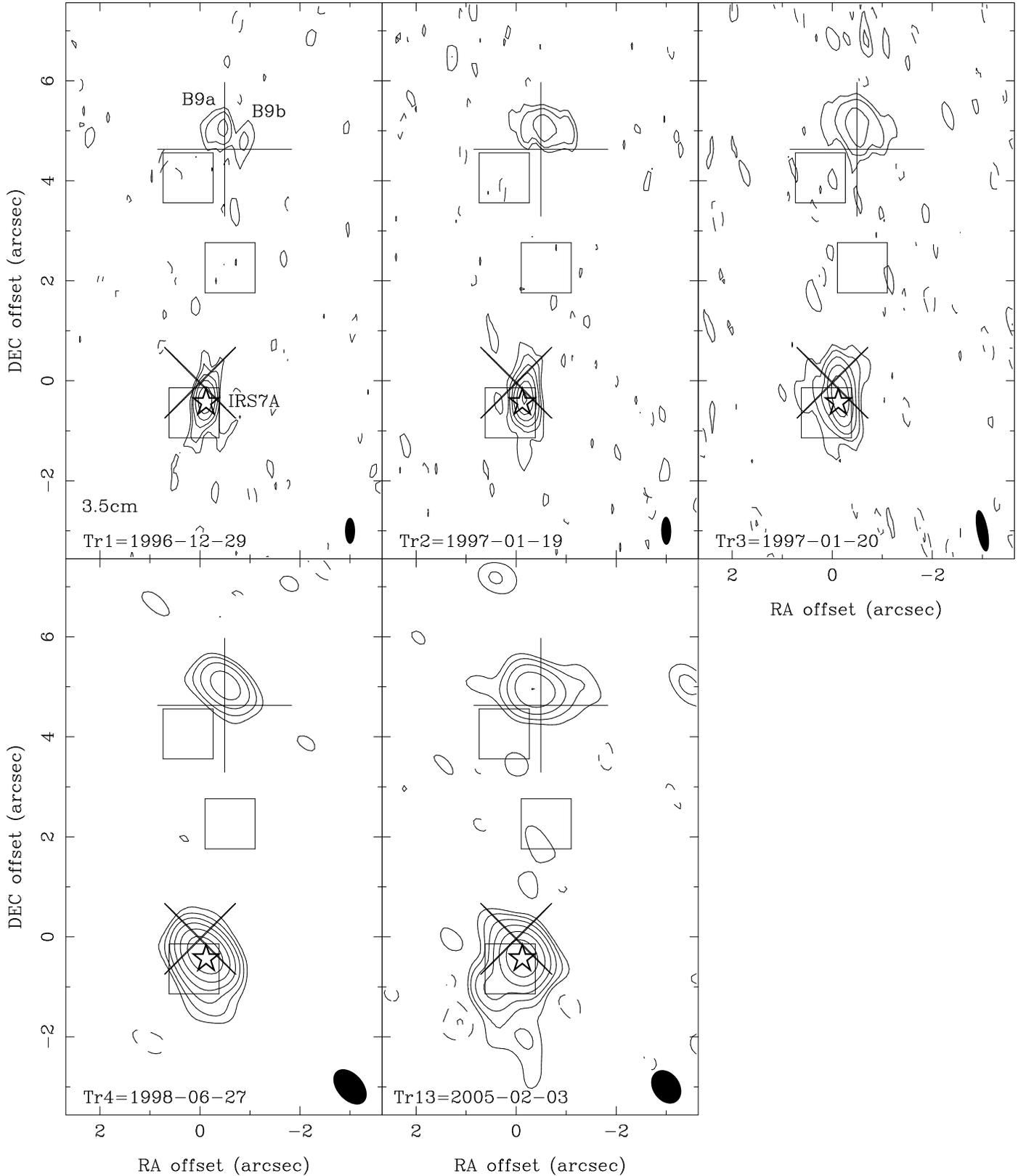

\ploteps{IRS7A.Xuni.eps}{185mm}
\caption{\small\baselineskip=0.825\baselineskip
Uniform-weight maps of the 3.5 cm continuum toward IRS 7A and B 9.
The contour levels are 1, 2, 4, 8, 16, and 32 times 0.06 m\Jypb.
A position correction was applied to the Tr 2 and 3 maps (see \S~3.2).
{\it Plus sign}:
The 1.1 mm continuum source SMA 2 (Groppi et al. 2007).
{\it Squares}:
The 7 mm continuum sources CT 2, 3, and 4 (Choi \& Tatematsu 2004).
{\it Cross}:
The X-ray source CXOU J190155.3--365722 (Hamaguchi et al. 2005).
{\it Star symbol}:
The mid-IR source IRAC 5 (Groppi et al. 2007).}
\end{figure*}

%% file: fig7.tex
\begin{figure}
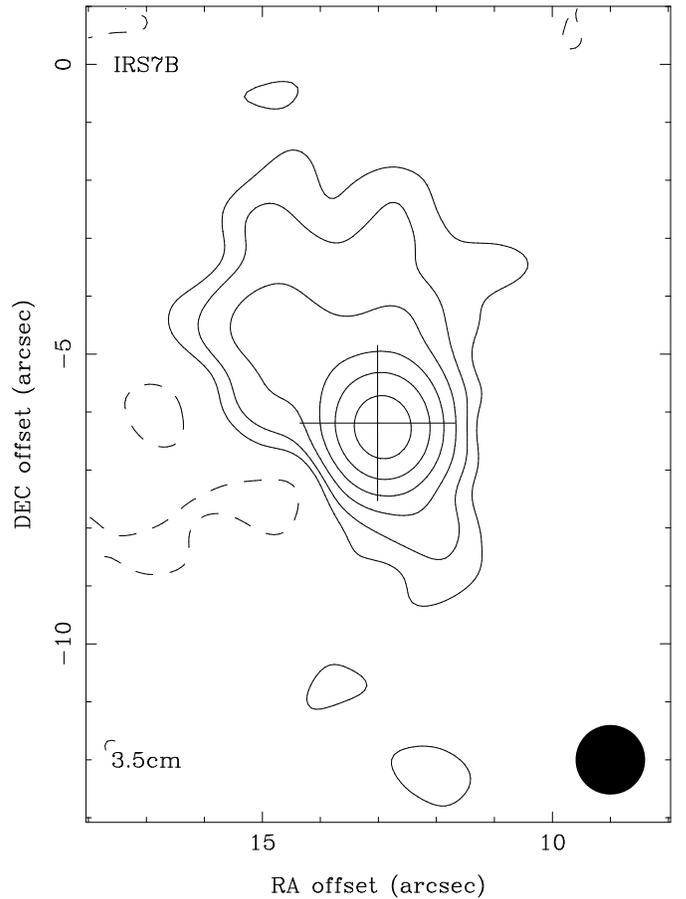

\ploteps{IRS7B.Xavg.eps}{88mm}
\caption{\small\baselineskip=0.825\baselineskip
Map of the 3.5 cm continuum toward the IRS 7B region
from high-resolution data sets: Tr 1, 2, 3, 4, and 13.
See Fig. 4 for details.
The contour levels are 1, 2, 4, 8, 16, and 32 times 0.025 m\Jypb.
{\it Plus sign}:
The 1.1 mm continuum source SMA 1 (Groppi et al. 2007).}
\end{figure}

%% file: fig8.tex
\begin{figure}
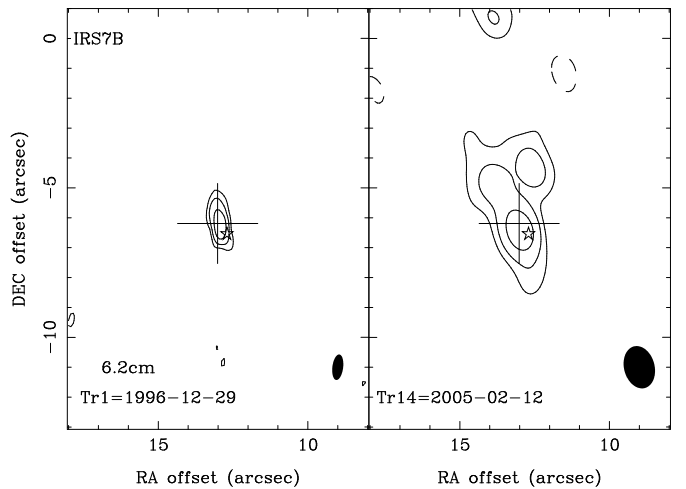

\ploteps{IRS7B.Cuni.eps}{88mm}
\caption{\small\baselineskip=0.825\baselineskip
Uniform-weight maps of the 6.2 cm continuum toward the IRS 7B region.
The contour levels are 1, 2, and 4 times 0.10 m\Jypb.
{\it Plus sign}:
The 1.1 mm continuum source SMA 1 (Groppi et al. 2007).
{\it Star symbol}:
The mid-IR source IRAC 4 (Groppi et al. 2007).}
\end{figure}

%% file: fig9.tex
\begin{figure*}
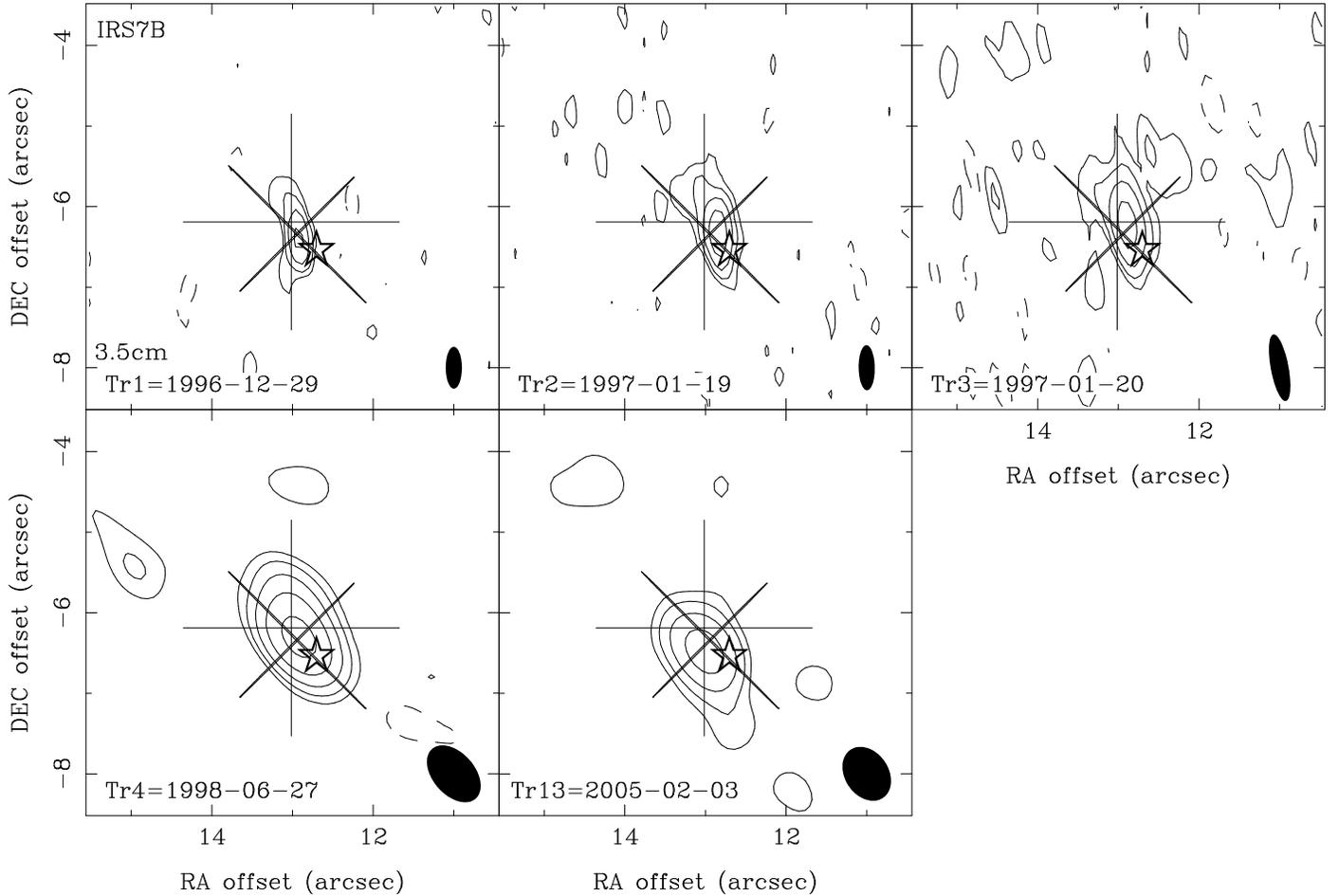

\ploteps{IRS7B.Xuni.eps}{185mm}
\caption{\small\baselineskip=0.825\baselineskip
Uniform-weight maps of the 3.5 cm continuum toward IRS 7B.
The contour levels are 1, 2, 4, 8, and 16 times 0.06 m\Jypb.
{\it Plus sign}:
The 1.1 mm continuum source SMA 1 (Groppi et al. 2007).
{\it Cross}:
The X-ray source CXOU J190156.4--365728 (Hamaguchi et al. 2005).
{\it Star symbol}:
The mid-IR source IRAC 4 (Groppi et al. 2007).}
\end{figure*}

%% file: fig10.tex
\begin{figure}
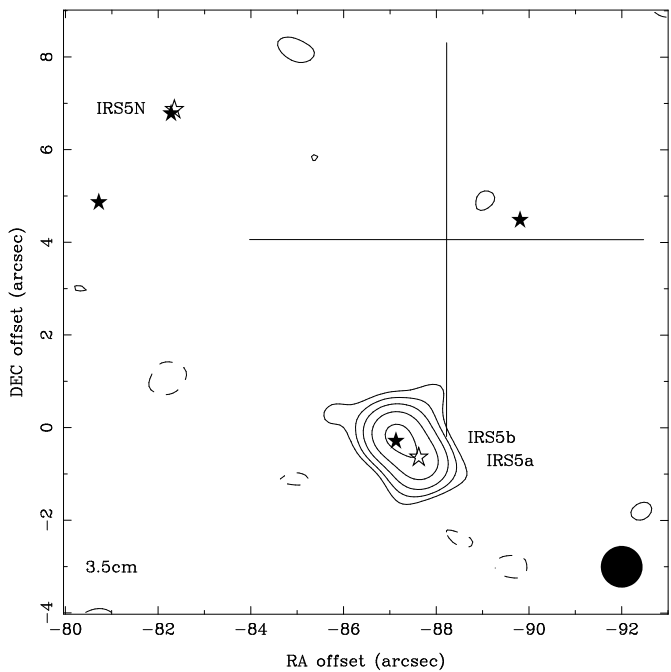

\ploteps{IRS5.Xavg.eps}{88mm}
\caption{\small\baselineskip=0.825\baselineskip
Map of the 3.5 cm continuum toward the IRS 5 region
from high-resolution data sets: Tr 1, 2, 3, 4, and 13.
A natural-weight image was made from the Tr 1 data set,
and a uniform-weight image was made from each of the other data sets.
Each of these images was corrected for the primary beam response
and convolved to an angular resolution of 0.9$''$
(shown in the bottom right corner),
and then all the five images were averaged.
The contour levels are 1, 2, 4, 8, and 16 times 0.04 m\Jypb.
{\it Plus sign}:
The submillimeter continuum source SMM 4 (Groppi et al. 2007).
{\it Open star symbols}:
The mid-IR sources IRAC 9 (IRS 5N) and IRAC 10 (Groppi et al. 2007).
{\it Filled star symbols}:
Near-IR sources listed in the 2MASS Point Source Catalogue.}
\end{figure}

%% file: fig11.tex
\begin{figure}
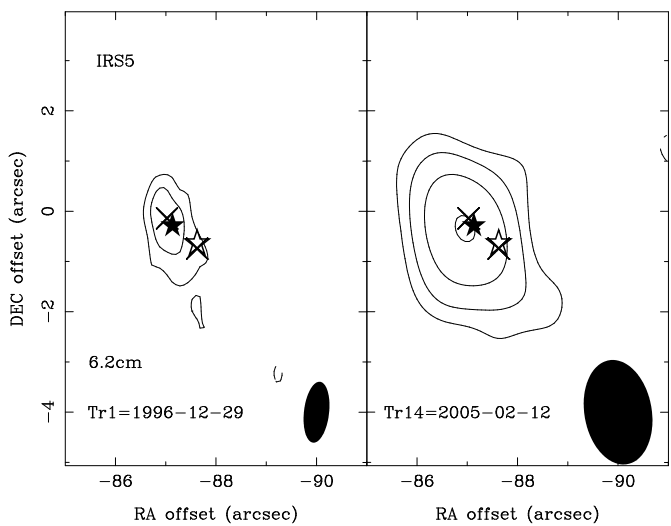

\ploteps{IRS5.C.eps}{88mm}
\caption{\small\baselineskip=0.825\baselineskip
Natural-weight maps of the 6.2 cm continuum toward the IRS 5 region.
The contour levels are 1, 2, 4, and 8 times 0.07 m\Jypb.
{\it Star symbols}:
The same as those in Fig. 10.
{\it Crosses}:
Sources detected in X-rays (Hamaguchi et al. 2008).}
\end{figure}

%% file: fig12.tex
\begin{figure*}
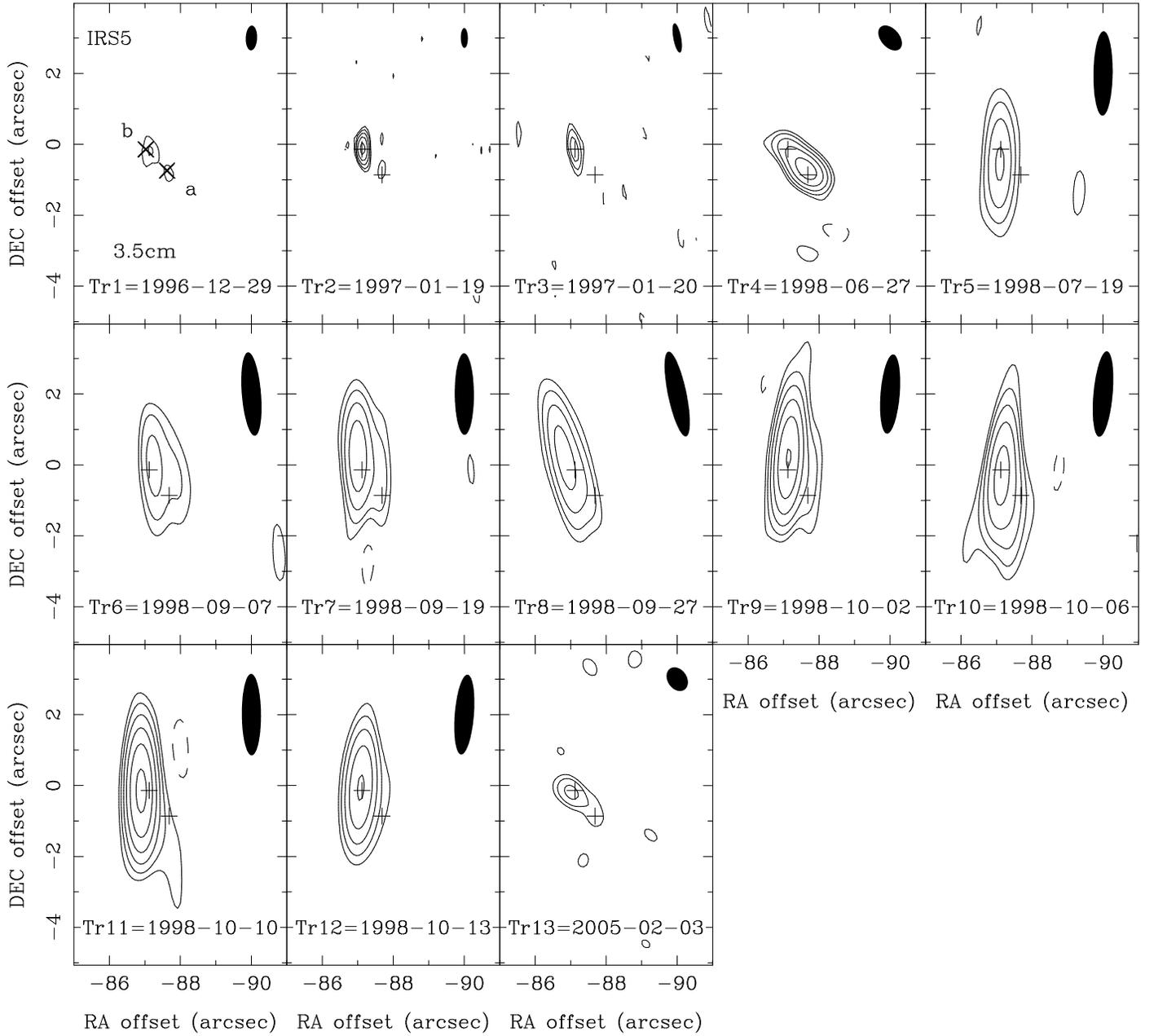

\ploteps{IRS5.Xuni.eps}{185mm}
\caption{\small\baselineskip=0.825\baselineskip
Maps of the 3.5 cm continuum toward IRS 5.
The first map (Tr 1) was made with a natural weighting,
and the others were made with a uniform weighting.
The contour levels are 1, 2, 4, 8, 16, and 32 times 0.08 m\Jypb.
{\it Crosses}:
Sources detected in X-rays (Hamaguchi et al. 2008).
{\it Plus signs}:
Peak positions of the 3.5 cm sources in the Tr 1 map ({\it first panel}).}
\end{figure*}

%% file: tab5.tex
\begin{deluxetable}{p{16mm}ccccc}
\tabletypesize{\small}
\tablecaption{Flux Densities of the IRS 5 Binary Components at 3.5 cm}%
\tablewidth{0pt}
\tablehead{
& \multicolumn{2}{c}{IRS 5a} && \multicolumn{2}{c}{IRS 5b} \\
\cline{2-3} \cline{5-6}
\colhead{\sc Track} & \colhead{Peak} & \colhead{Total}
&& \colhead{Peak} & \colhead{Total} }%
\startdata
Tr 1\dotfill  & 0.11 $\pm$ 0.02 & 0.12 $\pm$ 0.02   && 0.17 $\pm$ 0.02 & 0.25 $\pm$ 0.02 \\
Tr 2\dotfill  & 0.15 $\pm$ 0.03 & 0.15 $\pm$ 0.03   && 1.63 $\pm$ 0.03 & 1.82 $\pm$ 0.04 \\
Tr 3\dotfill  & \nodata         & \nodata           && 0.51 $\pm$ 0.03 & 0.58 $\pm$ 0.04 \\
Tr 4\dotfill  & 1.93 $\pm$ 0.03 & 2.04 $\pm$ 0.05   && \nodata         & 0.53 $\pm$ 0.04 \\
Tr 5\dotfill  & \nodata         & \nodata           && 0.72 $\pm$ 0.03 & 0.75 $\pm$ 0.04 \\
Tr 6\dotfill  & \nodata         & 0.16 $\pm$ 0.03   && 0.45 $\pm$ 0.03 & 0.45 $\pm$ 0.04 \\
Tr 7\dotfill  & \nodata         & 0.28 $\pm$ 0.04   && 1.10 $\pm$ 0.02 & 1.05 $\pm$ 0.04 \\
Tr 8\dotfill  & \nodata         & 0.26 $\pm$ 0.05   && 0.94 $\pm$ 0.02 & 0.91 $\pm$ 0.03 \\
Tr 9\dotfill  & \nodata         & 0.27 $\pm$ 0.04   && 2.67 $\pm$ 0.02 & 2.67 $\pm$ 0.04 \\
Tr 10\dotfill & \nodata         & 0.41 $\pm$ 0.03   && 1.77 $\pm$ 0.03 & 1.83 $\pm$ 0.04 \\
Tr 11\dotfill & \nodata         & 0.22 $\pm$ 0.04   && 3.11 $\pm$ 0.02 & 3.14 $\pm$ 0.05 \\
Tr 12\dotfill & \nodata         & 0.08 $\pm$ 0.04   && 1.38 $\pm$ 0.02 & 1.43 $\pm$ 0.04 \\
Tr 13\dotfill & 0.16 $\pm$ 0.03 & 0.16 $\pm$ 0.03   && 0.41 $\pm$ 0.03 & 0.41 $\pm$ 0.03 \\
\enddata\\
\tablecomments{Flux densities are corrected for the primary beam response,
               and the unit is mJy (beam$^{-1}$).
               The Tr 1 fluxes are from the natural-weight map,
               and the others are from the uniform-weight maps.
               For Tr 1, 2, 3, and 13,
               IRS 5a and 5b are relatively well-separated in the image,
               and each total flux was measured in a box around each source.
               For Tr 4--12,
               the binary components were not clearly separated,
               and the total fluxes were measured by fitting each image
               using a sum of two point-like sources.
               In the fitting process,
               the position offset between the two sources was fixed
               using the value measured in the Tr 1 map.}%
\end{deluxetable}

%% file: fig13.tex
\begin{figure}
\ploteps{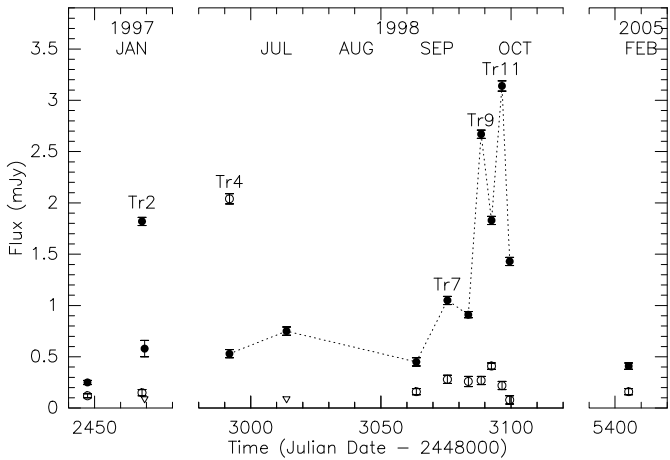}{88mm}
\caption{\small\baselineskip=0.825\baselineskip
Light curves of IRS 5a/b at 3.5 cm.
{\it Open circles}:
Flux densities of IRS 5a.
{\it Open triangles}:
Flux upper limits (3$\sigma$) for IRS 5a (Tr 3/5).
{\it Filled circles}:
Flux densities of IRS 5b.}
\end{figure}

%% file: fig14.tex
\begin{figure*}
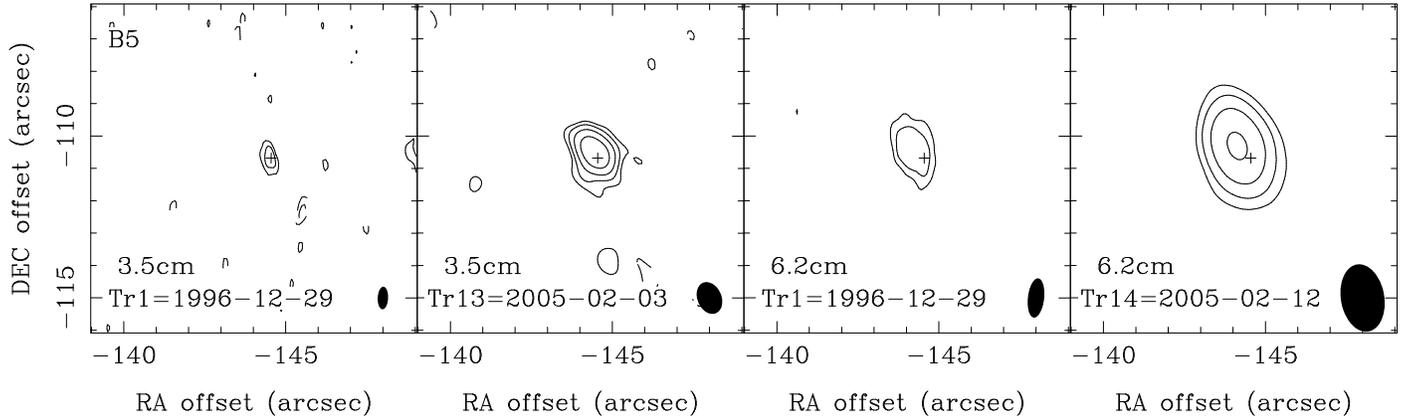

\ploteps{B5.XC.eps}{185mm}
\caption{\small\baselineskip=0.825\baselineskip
Natural-weight maps of the 3.5 cm ({\it left two panels})
and 6.2 cm ({\it right two panels}) continuum toward B 5.
The contour levels are 1, 2, 4, and 8 times 0.12 m\Jypb.
{\it Plus sign}:
The 3.5 cm continuum peak position of the Tr 1 map.}
\end{figure*}

%% file: fig15.tex
\begin{figure*}
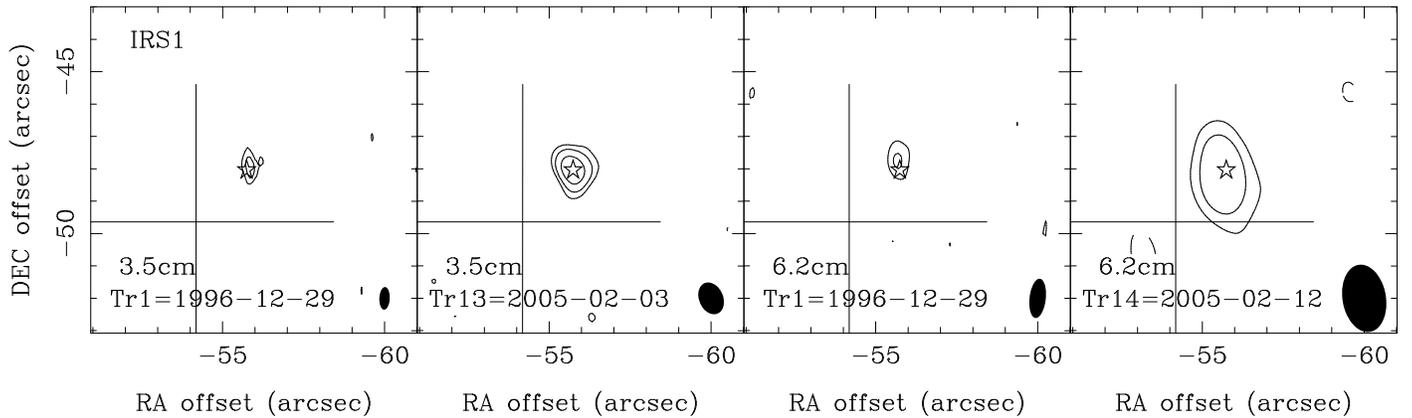

\ploteps{IRS1.XC.eps}{185mm}
\caption{\small\baselineskip=0.825\baselineskip
Natural-weight maps of the 3.5 cm ({\it left two panels})
and 6.2 cm ({\it right two panels}) continuum toward IRS 1.
The contour levels are 1, 2, and 4 times 0.07 m\Jypb.
{\it Plus sign}:
The submillimeter continuum source SMM 3 (Groppi et al. 2007).
{\it Star symbol}:
The mid-IR source IRAC 7 (Groppi et al. 2007).}
\end{figure*}

%% file: fig16.tex
\begin{figure*}
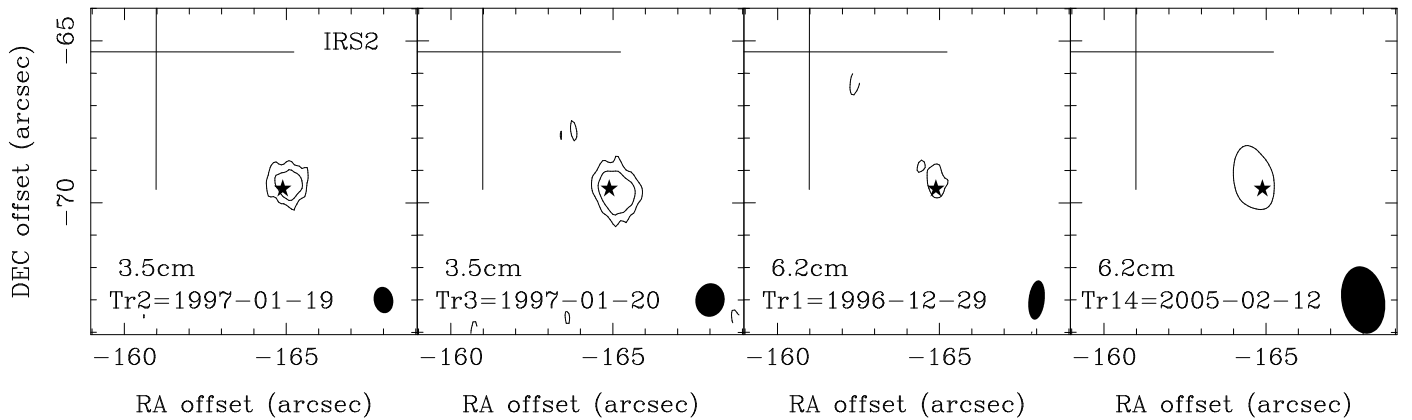

\ploteps{IRS2.XC.eps}{185mm}
\caption{\small\baselineskip=0.825\baselineskip
Natural-weight maps of the 3.5 cm ({\it left two panels})
and 6.2 cm ({\it right two panels}) continuum toward IRS 2.
The contour levels are 1 and 2 times 0.09 m\Jypb.
{\it Plus sign}:
The submillimeter continuum source SMM 5 (Nutter et al. 2005).
{\it Star symbol}:
Near-IR source listed in the 2MASS Point Source Catalogue.}
\end{figure*}

%% file: fig17.tex
\begin{figure*}
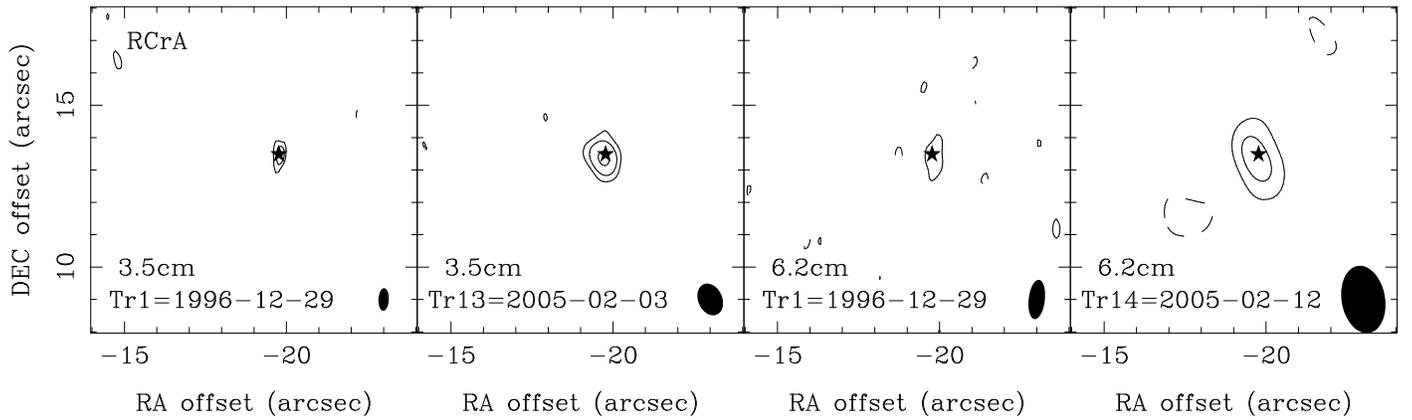

\ploteps{RCrA.XC.eps}{185mm}
\caption{\small\baselineskip=0.825\baselineskip
Natural-weight maps of the 3.5 cm ({\it left two panels})
and 6.2 cm ({\it right two panels}) continuum toward R CrA.
The contour levels are 1, 2, and 4 times 0.06 m\Jypb.
{\it Star symbol}:
Near-IR source listed in the 2MASS Point Source Catalogue.}
\end{figure*}

%% file: fig18.tex
\begin{figure*}
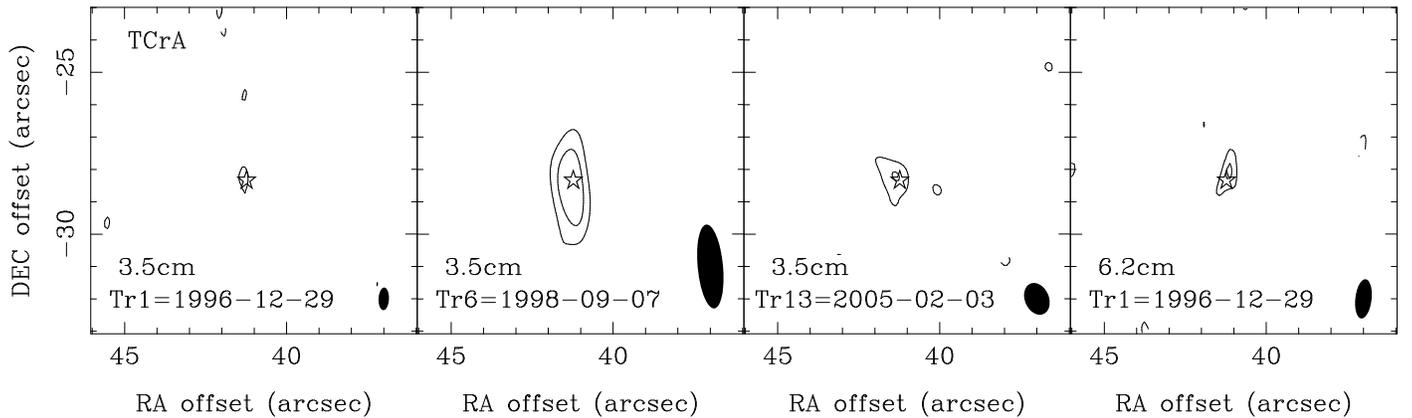

\ploteps{TCrA.XC.eps}{185mm}
\caption{\small\baselineskip=0.825\baselineskip
Natural-weight maps of the 3.5 cm ({\it left three panels})
and 6.2 cm ({\it rightmost panel}) continuum toward T CrA.
The contour levels are 1 and 2 times 0.06 m\Jypb.
{\it Star symbol}:
The mid-IR source IRAC 1 (Groppi et al. 2007).}
\end{figure*}

%% file: tab6.tex
\begin{deluxetable}{p{16mm}ccccc}
\tabletypesize{\small}
\tablecaption{Flux Densities of T CrA and WMB 55 at 3.5 cm}%
\tablewidth{0pt}
\tablehead{
& \multicolumn{2}{c}{\sc T CrA} && \multicolumn{2}{c}{\sc WMB 55} \\
\cline{2-3} \cline{5-6}
\colhead{\sc Track}
& \colhead{Peak} & \colhead{Total} && \colhead{Peak} & \colhead{Total} }%
\startdata
Tr 1\dotfill  & 0.10 $\pm$ 0.02 & 0.10 $\pm$ 0.02 && \nodata         & \nodata \\
Tr 2\dotfill  & 0.11 $\pm$ 0.02 & 0.21 $\pm$ 0.03 && \nodata         & \nodata \\
Tr 3\dotfill  & 0.11 $\pm$ 0.02 & 0.11 $\pm$ 0.02 && \nodata         & \nodata \\
Tr 4\dotfill  & 0.10 $\pm$ 0.02 & 0.10 $\pm$ 0.02 && 0.13 $\pm$ 0.02 & 0.13 $\pm$ 0.02 \\
Tr 5\dotfill  & \nodata         & \nodata         && \nodata         & \nodata \\
Tr 6\dotfill  & 0.19 $\pm$ 0.02 & 0.20 $\pm$ 0.02 && \nodata         & \nodata \\
Tr 7\dotfill  & 0.11 $\pm$ 0.03 & 0.13 $\pm$ 0.03 && \nodata         & \nodata \\
Tr 8\dotfill  & 0.11 $\pm$ 0.03 & 0.13 $\pm$ 0.03 && \nodata         & \nodata \\
Tr 9\dotfill  & 0.16 $\pm$ 0.03 & 0.16 $\pm$ 0.03 && \nodata         & \nodata \\
Tr 10\dotfill & 0.08 $\pm$ 0.02 & 0.08 $\pm$ 0.02 && \nodata         & \nodata \\
Tr 11\dotfill & 0.16 $\pm$ 0.03 & 0.16 $\pm$ 0.03 && 0.15 $\pm$ 0.03 & 0.15 $\pm$ 0.03 \\
Tr 12\dotfill & \nodata         & \nodata         && 0.18 $\pm$ 0.03 & 0.20 $\pm$ 0.04 \\
Tr 13\dotfill & 0.13 $\pm$ 0.02 & 0.17 $\pm$ 0.02 && 0.15 $\pm$ 0.02 & 0.17 $\pm$ 0.03 \\
\enddata\\
\tablecomments{Flux densities are corrected for the primary beam response,
               and the unit is mJy (beam$^{-1}$).
               They are from the natural-weight maps.}%
\end{deluxetable}

%% file: fig19.tex
\begin{figure*}
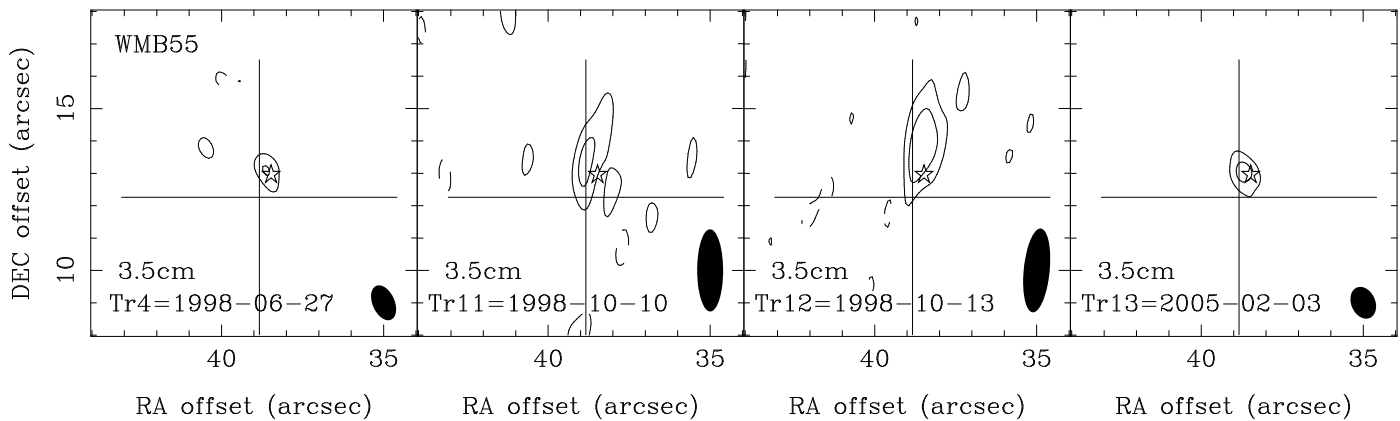

\ploteps{WMB55.XC.eps}{185mm}
\caption{\small\baselineskip=0.825\baselineskip
Natural-weight maps of the 3.5 cm continuum toward WMB 55.
The contour levels are 1 and 2 times 0.06 m\Jypb.
{\it Plus sign}:
The submillimeter continuum source SMM 2 (Groppi et al. 2007).
{\it Star symbol}:
The mid-IR source IRAC 3 (Groppi et al. 2007).}
\end{figure*}

%% file: fig20.tex
\begin{figure}[t]
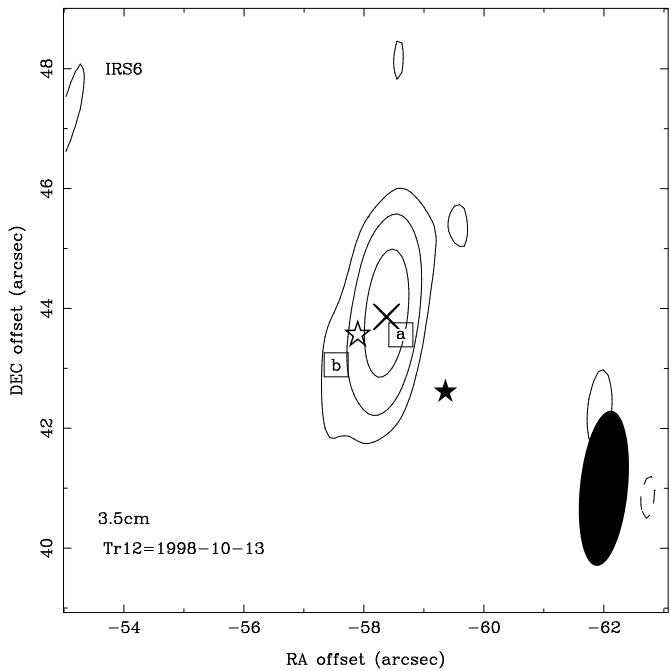

\ploteps{IRS6.X.eps}{88mm}
\caption{\small\baselineskip=0.825\baselineskip
Natural-weight map of the 3.5 cm continuum toward IRS 6.
The contour levels are 1, 2, and 4 times 0.06 m\Jypb.
{\it Cross}:
X-ray source (Hamaguchi et al. 2008).
{\it Squares}:
Near-IR sources in Fig. 1 of Nisini et al. (2005).
A position correction was applied
using the offset determined with IRS 5a/b (see text).
The binary separation is $\sim$1.2$''$.
{\it Open star symbol}:
The mid-IR source IRAC 8 (Groppi et al. 2007).
{\it Filled star symbol}:
Near-IR source listed in the 2MASS Point Source Catalogue.
Note that this 2MASS source is not IRS 6 but a nebulosity.
IRS 6 itself is listed in the Point Source Reject Table
and coincides with IRAC 8 within $\sim$0.2$''$.}
\end{figure}

%% file: ms.bbl
\begin{references}
\reference{} Allen, L. E., et al. 2004, ApJS, 154, 363
\reference{} Anderson, I. M., Harju, J., Knee, L. B. G.,
             \& Haikala, L. K. 1997, A\&A, 321, 575
\reference{} Anglada, G., Villuendas, E., Estalella, R.,
             Beltr{\'a}n, M. T., Rodr{\'\i}guez, L. F., Torrelles, J. M.,
             \& Curiel, S. 1998, AJ, 116, 2953
\reference{} Bally, J., Feigelson, E., \& Reipurth, B. 2003, ApJ, 584, 843
\reference{} Brown, A. 1987, ApJ, 322, L31
\reference{} Chen, H., Myers, P. C., Ladd, E. F., \& Wood, D. O. S. 1995,
             ApJ, 445, 377
\reference{} Chen, W. P., \& Graham, J. A. 1993, ApJ, 409, 319
\reference{} Chini, R., et al. 2003, A\&A, 409, 235
\reference{} Choi, M., \& Tatematsu, K. 2004, ApJ, 600, L55
\reference{} Choi, M., Tatematsu, K., Hamaguchi, K., \& Lee, J.-E. 2008,
             ApJ, submitted (Paper II)
\reference{} Crusius-W{\"a}tzel, A. R. 1990, ApJ, 361, L49
\reference{} Dulk, G. A. 1985, ARA\&A, 23, 169
\reference{} Favata, F., Fridlund, C. V. M., Micela, G., Sciortino, S.,
             \& Kaas, A. A. 2002, A\&A, 386, 204
\reference{} Feigelson, E. D., Carkner, L., \& Wilking, B. A. 1998,
             ApJ, 494, L215
\reference{} Feigelson, E. D., \& Montmerle, T. 1999, ARA\&A, 37, 363
\reference{} Forbrich, J., \& Preibisch, T. 2007, A\&A, 475, 959
\reference{} Forbrich, J., Preibisch, T., \& Menten, K. M. 2006,
             A\&A, 446, 155
\reference{} Forbrich, J. et al. 2007, A\&A, 464, 1003
\reference{} Garay, G., Brooks, K. J., Mardones, D., \& Norris, R. P. 2003,
             ApJ, 587, 739
\reference{} Groppi, C. E., Hunter, T. R., Blundell, R., \& Sandell, G. 2007,
             ApJ, 670, 489
\reference{} Groppi, C. E., Kulesa, C., Walker, C., \& Martin, C. L. 2004,
             ApJ, 612, 946
\reference{} Hamaguchi, K., Choi, M., Corcoran, M. F., Choi, C.-S.,
             Tatematsu, K., \& Petre, R. 2008, ApJ, in press
\reference{} Hamaguchi, K., Corcoran, M. F., Petre, R., White, N. E.,
             Stelzer, B., Nedachi, K., Kobayashi, N.,
             \& Tokunaga, A. T. 2005, ApJ, 623, 291
\reference{} Harju, J., Haikala, L. K., Mattila, K., Mauersberger, R.,
             Booth, R. S., \& Nordh, H. L. 1993, A\&A, 278, 569
\reference{} Hartigan, P., \& Graham, J. A. 1987, AJ, 93, 913
\reference{} Henning, Th., Launhardt, R., Steinacker, J., \& Thamm, E. 1994,
             A\&A, 291, 546
\reference{} Henriksen, R. N., Ptuskin, V. S., \& Mirabel, I. F. 1991,
             A\&A, 248, 221
\reference{} Knude, J., \& H{\o}g, E. 1998, A\&A, 338, 897
\reference{} Koyama, K., Hamaguchi, K., Ueno, S., Kobayashi, N.,
             \& Feigelson, E. D. 1996, PASJ, 48, L87
\reference{} Lee, J.-E., et al. 2006, ApJ, 648, 491
\reference{} Loinard, L., Rodr{\'\i}guez, L. F., D'Alessio, P.,
             Rodr{\'\i}guez, M. I., \& Gonz{\'a}lez, R. F. 2007, ApJ, 657, 916
\reference{} Massi, M., Forbrich, J., Menten, K. M.,
             Torricelli-Ciamponi, G., Neidh{\"o}fer, J., Leurini, S.,
             \& Bertoldi, F. 2006, A\&A, 453, 959
\reference{} Massi, M., et al. 2008, A\&A, 480, 489
\reference{} Miettinen, O., Kontinen, S., Harju, J., \& Higdon, J. L. 2008,
             A\&A, in press
\reference{} Montmerle, T., Grosso, N., Tsuboi, Y., \& Koyama, K. 2000,
             ApJ, 532, 1097
\reference{} Mutel, R. L., Lestrade, J. F., Preston, R. A.,
             \& Phillips, R. B. 1985, ApJ, 289, 262
\reference{} Neuh{\"a}user, R., et al. 2000, A\&AS, 146, 323
\reference{} Nisini, B., Antoniucci, S., Giannini, T.,
             \& Lorenzetti, D. 2005, A\&A, 429, 543
\reference{} Nutter, D. J., Ward-Thompson, D., \& Andr{\'e}, P. 2005,
             MNRAS, 357, 975
\reference{} Phillips, R. B., Lonsdale, C. J., \& Feigelson, E. D. 1991,
             ApJ, 382, 261
\reference{} Pravdo, S. H., Feigelson, E. D., Garmire, G., Maeda, Y.,
             Tsuboi, Y., \& Bally, J. 2001, Nature, 413, 708
\reference{} Reynolds, S. P. 1986, ApJ, 304, 713
\reference{} Rodr{\'\i}guez, L. F., Anglada, G., \& Curiel, S. 1999,
             ApJS, 125, 427
\reference{} Rodr{\'\i}guez, L. F., Curiel, S., Moran, J. M., Mirabel, I. F.,
             Roth, M., \& Garay, G. 1989, ApJ, 346, L85
\reference{} Strom, K. M., Strom, S. E., \& Grasdalen, G. L. 1974,
             ApJ, 187, 83
\reference{} Suters, M., Stewart, R. T., Brown, A., \& Zealey, W. 1996,
             AJ, 111, 320
\reference{} Takami, M., Bailey, J., \& Chrysostomou, A. 2003, A\&A, 397, 675
\reference{} Taylor, K. N. R., \& Storey, J. W. V. 1984, MNRAS, 209, 5P
\reference{} Tsujimoto, M., Koyama, K., Kobayashi, N., Saito, M.,
             Tsuboi, Y., \& Chandler, C. J. 2004, PASJ, 56, 341
\reference{} van den Ancker, M. E. 1999, Ph.D. thesis, Univ. Amsterdam
\reference{} Wang, H., Mundt, R., Henning, T., \& Apai, D. 2004,
             ApJ, 617, 1191
\reference{} Wilking, B. A., Greene, T. P., Lada, C. J., Meyer, M. R.,
             \& Young, E. T. 1992, ApJ, 397, 520
\reference{} Wilking, B. A., McCaughrean, M. J., Burton, M. G., Giblin, T.,
             Rayner, J. T., \& Zinnecker, H. 1997, AJ, 114, 2029
\end{references}
